\documentclass[acmsmall,screen]{acmart}

\makeatletter
\def\@ACM@copyright@check@cc{%
}
\makeatother

\usepackage[a-1b]{pdfx}

\usepackage{pdfprivacy}
\usepackage{textpos}

\usepackage[
type={CC},
modifier={by},
version={4.0},
imagewidth=7em,
]{doclicense}

\AtBeginDocument{%
  \providecommand\BibTeX{{%
    \normalfont B\kern-0.5em{\scshape i\kern-0.25em b}\kern-0.8em\TeX}}}

\setcopyright{cc}
\acmJournal{TOPS}
\acmYear{2022} \acmVolume{26} \acmNumber{1} \acmArticle{6} \acmMonth{11} 
\acmPrice{}\acmDOI{10.1145/3546069}

\usepackage{float} %

\usepackage{graphicx}

\usepackage[caption=false]{subfig}

\usepackage{wasysym}

\usepackage[inline]{enumitem}

\usepackage{color}

\setenumerate{label=(\roman*), leftmargin=2em}

\newcommand{\boldparagraph}[1]{\vspace{1em}\noindent\textbf{#1.}}

\newcommand{\censorhidden}[1]{}
\newcommand{\censorchange}[2]{#2}
\newcommand{\censor}[1]{\textit{\textless withheld during blind review \textgreater}}

\renewcommand{\censorhidden}[1]{#1}
\renewcommand{\censorchange}[2]{#1}
\renewcommand{\censor}[1]{#1}

\usepackage{soul}
\usepackage{xcolor}

\definecolor{hlcolor}{RGB}{214, 239, 255}\sethlcolor{hlcolor}%
\newcommand{\revisionadd}[1]{\ctext[RGB]{214, 239, 255}{#1}}
\newcommand{\revisionaddimage}[1]{
	\setlength{\fboxsep}{0pt}%
	\setlength{\fboxrule}{1pt}%
	\fcolorbox{hlcolor}{hlcolor}{
	\hspace{-1.5mm}#1}
}

\renewcommand{\revisionadd}[1]{#1}
\renewcommand{\revisionaddimage}[1]{#1}

\usepackage{tcolorbox}

\newcommand{\findingsbox}[1]{\vspace{-0.5em}\begin{tcolorbox}[title=Main Findings,boxsep=0.3mm,left=2mm,right=2mm,bottom=1mm,top=1mm]
		\revisionadd{#1}\end{tcolorbox}} \begin{document}

\title[Pump Up Password Security! Evaluating and Enhancing RBA on a Real-World Large-Scale Online Service]{Pump Up Password Security! Evaluating and Enhancing Risk-Based Authentication on a Real-World Large-Scale Online Service}

\author{Stephan Wiefling}
\email{stephan.wiefling@rub.de}
\orcid{0000-0001-7917-6065}
\affiliation{%
  \institution{H-BRS University of Applied Sciences}
  \city{Sankt Augustin}
  \country{Germany}
}
\affiliation{%
	\institution{Ruhr University Bochum}
	\city{Bochum}
	\country{Germany}
}

\author{Paul René Jørgensen}
\orcid{0000-0003-3806-714X}
\author{Sigurd Thunem}
\orcid{0000-0001-7569-8501}
\affiliation{%
  \institution{Telenor Digital}
  \city{Fornebu}
  \country{Norway}
}

\author{Luigi Lo Iacono}
\email{luigi.lo_iacono@h-brs.de}
\orcid{0000-0002-7863-0622}
\affiliation{%
	\institution{H-BRS University of Applied Sciences}
	\city{Sankt Augustin}
	\country{Germany}
}

\renewcommand{\shortauthors}{S. Wiefling et al.}

\begin{abstract}
Risk-based authentication (RBA) aims to protect users against attacks involving stolen passwords. RBA monitors features during login, and requests re-authentication when feature values widely differ from previously observed ones. It is recommended by various national security organizations, and users perceive it more usable and equally secure than equivalent two-factor authentication. Despite that, RBA is still only used by very few online services. Reasons for this include a lack of validated open resources on RBA properties, implementation, and configuration. This effectively hinders the RBA research, development, and adoption progress.

To close this gap, we provide the first long-term RBA analysis on a real-world large-scale online service. We collected feature data of 3.3 million users and 31.3 million login attempts over more than one year. Based on the data, we provide (i) studies on RBA's real-world characteristics, and its configurations and enhancements to balance usability, security, and privacy, (ii) a machine learning based RBA parameter optimization method to support administrators finding an optimal configuration for their own use case scenario, (iii) an evaluation of the round-trip time feature's potential to replace the IP address for enhanced user privacy, and (iv) a synthesized RBA data set to reproduce this research and to foster future RBA research.
Our results provide insights on selecting an optimized RBA configuration so that users profit from RBA after just a few logins. The open data set enables researchers to study, test, and improve RBA for widespread deployment in the wild. \end{abstract}



\maketitle

\section{Introduction}

Despite decades of efforts to replace passwords with more secure alternatives~\cite{bonneau_quest_2012}, they are still the main authentication method for the majority of online services~\cite{quermann_state_2018}. The recent rise of data breaches further increased the threats to password-based authentication~\cite{haveibeenpwned_pwned_2020,thomas_data_2017,mayer_now_2021}. Credential stuffing and password spraying attacks automatically enter leaked login credentials (username and password) on other websites, in the hope that users re-used them~\cite{haber_attack_2020,das_tangled_2014}. Such attacks can be scaled with little effort and promise financial returns, which is why they are very popular today~\cite{overson_state_2020,akamai_loyalty_2020}. Targeted password guessing methods using machine learning (ML) are less scalable, but also very effective~\cite{wang_targeted_2016,pal_beyond_2019}. %
This increases the need for responsible online services to protect their users. %
Two-factor authentication (2FA)~\cite{petsas_two-factor_2015} is a commonly offered measure to increase account security, but users tend to refuse it outside of use cases involving sensitive and financial data, like, e.g., online banking~\cite{twitter_account_2021,newman_facebook_2021,milka_anatomy_2018,wiefling_more_2020,reese_usability_2019,dutson_dont_2019,redmiles_you_2017}. \revisionadd{For instance, the 2FA adoption rates are low on popular online services, e.g., 2.5\% on Twitter~{\cite{twitter_account_2021}} and $\approx$4\% on Facebook in 2021~{\cite{newman_facebook_2021}}. Google also has a low 2FA adoption rate~}{\cite{milka_anatomy_2018,petsas_two-factor_2015}}\revisionadd{, which is why they tried to force 2FA to 150M of their users, i.e., less than 10\%~{\cite{google_1.5_2018}}, by the end of 2021~{\cite{google_making_2021}}. The outcome of this experiment regarding user acceptance, however, is unclear to date. } %
To still protect users without enabled 2FA, and to increase the cost for attackers, risk-based authentication (RBA) is a solution that is preferred by users over equivalent 2FA in many use case scenarios~\cite{wiefling_more_2020}. It can be considered a scalable interim solution to strengthen password-based authentication, until more secure authentication methods are in place.
Therefore, government agencies like the National Institute of Standards and Technology (NIST, USA), the National Cyber Security Centre (NCSC, UK), and the Australian Cyber Security Centre~(ACSC, AU) recommend RBA to protect users against attacks involving stolen passwords~\cite{grassi_digital_2017,national_cyber_security_centre_cloud_2018,australian_australian_2021}. US federal agencies even have to establish RBA across their enterprise by presidential executive order~\cite{biden_jr_executive_2021}.

\subsection{Risk-Based Authentication (RBA)}
RBA estimates whether a login is legitimate or an account takeover attempt. This is done by monitoring and recording a set of features that are available in the login context. Potential features range from network (e.g., IP address), device or client (e.g., user agent string), to behavioral biometric information (e.g., login time)~\cite{wiefling_whats_2021}. Based on the feature values of the current login attempt and those of previous ones stored in a login history, RBA calculates a risk score related to the login attempt. Optionally, the risk score can also consider known attack data to identify suspicious feature values or potential users at risk (e.g., VIPs)~\cite{freeman_who_2016}. An access threshold typically classifies this risk score into low, medium, and high risk~\cite{wiefling_is_2019}. Based on the risk classification, the online service performs different actions~\cite{freeman_who_2016, wiefling_is_2019}. On a low risk (e.g., same device and location as always), access is granted without further service intervention. On a medium or higher risk (e.g., unusual device or location), RBA can ask for additional information to proof the claimed identity (e.g., verifying the email address), or even block access. %
The choice of the access threshold impacts RBA's overall usability and security~\cite{wiefling_whats_2021}. When the threshold is set too low, RBA requests re-authentication every login attempt, making the look and feel identical to classical 2FA. When set too high, RBA never requests re-authentication, but also does not provide any security benefit, as all attacks are classified as legitimate login attempts. Sensible RBA implementations range between these extremes. They only request re-authentication for a fraction of legitimate login attempts, but also protect users against a high number of attacks.

Some popular online services started using RBA~\cite{wiefling_is_2019,overson_state_2020,wiefling_whats_2021}. However, there is still a lack of open resources to support research and development. Only few research papers scientifically evaluated RBA algorithms and configurations, and these only used small data sets~\cite{freeman_who_2016, wiefling_whats_2021, wiefling_privacy_2021}. Besides that, there are no open source RBA solutions available, beyond a very basic and weak algorithm provided by the single sign-on (SSO) solution OpenAM~\cite{openam_adaptive_2016,wiefling_whats_2021}. More open resources can support online services to fully understand RBA and therefore enable a broad RBA deployment in practice.

\subsection{Research Questions and Contributions}

Following the governmental recommendations and obligations, more and more online services have to deploy RBA in a few months~\cite{biden_jr_executive_2021}. For real-world large-scale online services, however, RBA has not been widely understood to date. This is crucial, as even small configuration errors may largely impact usability and security for many users~\cite{wiefling_whats_2021}. %
\revisionadd{Previous studies already investigated RBA on a small online service~}{\cite{wiefling_whats_2021,wiefling_privacy_2021}}\revisionadd{, but they have limitations: They are limited to a small online service with a user base mostly located in one city. Therefore, it is unclear if the results on the RBA behavior hold for a globally distributed large user base.}

For these reasons, we provide the first in-depth analysis of RBA on a large-scale online service, including 3.3M users and 31.3M login attempts collected over more than one year. We evaluated RBA with four attacker models related to real-world attacks. The service\censorchange{, belonging to the telecommunications provider Telenor,}{} provided access to sensitive data, e.g., cloud storage and billing data. This made it a suitable RBA use case~\cite{wiefling_more_2020}. To investigate RBA on a large online service, we formulated and answered a set of research questions, which we outline below with our contributions. We answered the questions with the collected data. Two of the following research questions (RQ1a and RQ1c) replicate a related study~\cite{wiefling_whats_2021}, while the other research questions aim at enhancing our understanding of RBA on a large online service.

\newlist{RQLIST}{enumerate}{1}
\setlist[RQLIST]{label=\bfseries RQ\arabic*:, leftmargin=2.7em, parsep=0em, topsep=0.5em}

\newlist{RQ2LIST}{enumerate}{2}
\setlist[RQ2LIST]{label=\bgroup\bfseries \alph*)\egroup,leftmargin=1.3em, parsep=0em}

\vspace{1em}
\noindent
RBA's characteristics had previously been analyzed in a related study made on a small online service~\cite{wiefling_whats_2021}. %
A clear study limitation was that the data set was rather small and specific, and did not contain attack data. However, to investigate the broad applicability of the study's statement, it needs to be validated with a large user base. Therefore, we replicated this study, in order to validate whether the results also hold for a large online service. In so doing, we also analyzed how the RBA behavior depends on different login frequencies (e.g., daily to less frequent use). 
\begin{RQLIST}
	\item \begin{RQ2LIST}
		\item How often does RBA request re-authentication on a real-world large-scale online service?
		\item How does the number of re-authentication requests differ between users' login frequencies?
		\item How many user sessions need to be captured on a large-scale online service to achieve a stable and reliable setup?
		\item How do RBA characteristics differ between a large-scale and a small online service?
	\end{RQ2LIST}
\end{RQLIST}
\findingsbox{The results confirm previous findings~{\cite{wiefling_whats_2021}} that RBA rarely requests re-authentication in practice, even when blocking more than 99\% of attacks that used the victim's credentials and mimicked some of the legitimate login contexts. Going beyond the previous study, the results also show that the login frequencies significantly impact the number of re-authentication requests.}

\noindent
Other aspects of RBA have not yet been studied, e.g., due to a lack of available real-world, large-scale user login data~\cite{wiefling_privacy_2021,rivera_risk-based_2020}. These aspects included, e.g., how attack data~\cite{freeman_who_2016} and login history minimization~\cite{wiefling_privacy_2021} influence RBA's usability and security properties. Since our data set is large and contains attack data, we are feasible to analyze these aspects.
\begin{RQLIST}
  \setcounter{RQLISTi}{1}
	\item \begin{RQ2LIST}
		\item How does RBA perform when taking user-based attack data into account? 
		\item How does RBA perform when also taking feature-based attack data into account?
		\end{RQ2LIST}
	\item How long do we need to store login history data to maintain good RBA security and usability?
\end{RQLIST}
\findingsbox{The results suggest that attack data should only be used with caution or not at all. Login history minimization proved to be useful to increase privacy while balancing usability and security.}

\noindent
The current way of setting up RBA appears to be complicated in practice, as administrators need to set and test all parameters manually, which is error-prone and can largely impact usability and security~\cite{wiefling_whats_2021,rivera_risk-based_2020,freeman_who_2016}. Therefore, we introduce and tested a ML-based RBA parameter optimization that assists administrators to find an optimal RBA setup for their online service.
\begin{RQLIST}
  \setcounter{RQLISTi}{3}
	\item \begin{RQ2LIST}
		\item In what way can a classical RBA algorithm be enhanced with ML mechanisms to optimize the usability and security parameters?
		\item How do ML-based RBA optimizations perform compared to a classical algorithm?
	\end{RQ2LIST}
\end{RQLIST}
\findingsbox{The results show that the proposed optimization can provide a RBA configuration baseline to achieve good usability and security in short time.}

\noindent
The RBA performance of current algorithms likely degrades substantially with large global login history sizes, as is the case with large-scale online services~\cite{wiefling_whats_2021}. To tackle this problem, we also developed an algorithm optimization to speed up RBA authentication for large-scale online services.
\begin{RQLIST}
  \setcounter{RQLISTi}{4}
	\item What are performance implications of RBA on a large-scale online service and how can they be improved?
\end{RQLIST}
\findingsbox{The results identified non-optimized RBA as vulnerable to denial of service when users log in at similar times. Our optimizations using hash tables were able to shorten the authentication times with a 28x speedup.}

\noindent
Due to its hard-to-spoof properties, a server-originated round-trip time (RTT) feature~\cite{wiefling_whats_2021} showed potential to replace IP addresses as a privacy-enhancing alternative~\cite{wiefling_privacy_2021}. As this claim was not investigated and validated on a real user base so far, we evaluated the RTT's potential to verify countries and regions, and as an IP address replacement with our data set.
\begin{RQLIST}
  \setcounter{RQLISTi}{5}
	\item To what extent can the server-originated RTT be used as a privacy-enhancing alternative to the IP address feature?
\end{RQLIST}
\findingsbox{The results indicate that the RTT has strong potential as a drop-in replacement for the IP~address feature. It can identify legitimate users with good usability, security, and privacy while increasing the costs for attackers trying to spoof the IP geolocation, i.e., they need access to a device physically located in the target area instead of just using a VPN connection.}

\noindent
To address the current lack of high quality open data for RBA research, we further provide a synthesized RBA data set\footnote{\revisionadd{Available at} {\url{https://github.com/das-group/rba-dataset}}} (see Appendix~\ref{appendix:synthesized-data-set} for details). The synthesized data set resembles the statistical properties of our original data set that we had to delete for privacy reasons (see Section~\ref{sec:data-set}).

The results of our studies show that RBA's state of practice can significantly be improved. Its usability and security properties strongly depend on implementation and configuration details. Our research advances the RBA knowledge and development, and further supports the latter with the provided data set. It allows service owners to decide on RBA configurations for their online services. Administrators and developers get advice on selecting suitable RBA settings and optimizations to balance usability and security. Researchers get insights on RBA's behavior on a large online service.

\section{RBA Model Selection}\label{sec:rba-model-selection}

Different RBA models exist to derive a risk score for a login attempt. The first approach was published by Freeman et al.~\cite{freeman_who_2016} in 2016 and is based on a statistical relations. Other models, some more related to the area of implicit authentication, are based on more complex ML schemes such as siamese neural networks~\cite{acien_typenet_2020,rivera_risk-based_2020,deb_actions_2019} and one-class support vector machines (OCSVM)~\cite{misbahuddin_design_2017}. \revisionadd{Other relevant ML schemes for outlier detection include isolation forest, local outlier factor, and robust covariance~{\cite{scikit_novelty_2021}}. They have not been seen in RBA literature yet, but we found them relevant for our model selection.}
We used the RBA model of Freeman et al. since a preliminary evaluation showed it to be the most suitable for our data set. We outline the model selection in the following.

We aimed to focus on scalable and practical RBA solutions that can be used at large-scale online services. Like other popular online services observed in practice~\cite{wiefling_is_2019}, the online service collected categorical feature data, e.g., IP address and user agent. Two types of RBA models exist in literature that focus on categorical data~\cite{wiefling_whats_2021}. One type, the most published one (SIMPLE model), is very simple and only checks exact matches in the user's login history~\cite{steinegger_risk-based_2016,hurkala_architecture_2014,djosic_machine_2020,openam_adaptive_2016}. The other type, published by Freeman et al.~\cite{freeman_who_2016} and probably used by popular online services~\cite{wiefling_is_2019}, compares statistical feature relations between all users. This model outperformed the other models in a previous evaluation~\cite{wiefling_whats_2021}. \revisionadd{To provide a baseline comparison, we also repeated this evaluation on the large-scale online service (see Appendix~{\ref{appendix:simple-model}}). Both results confirm that the SIMPLE model can not block the same high amount of attackers as Freeman et al.'s model. The latter also allowed more fine-grained security configurations than the SIMPLE model. For instance, varying a tiny fraction of the access threshold greatly lowered the VPN attacker protection from 96\% to 53\% with the SIMPLE model on the large-scale online service. In contrast to that, the attacker protection only lowered for a tiny fraction with Freeman et al.'s model.} \revisionadd{On a geographically close user base, %
}it \revisionadd{also} blocked significantly more \revisionadd{targeted} attackers while making significantly fewer requests to re-authenticate~\cite{wiefling_whats_2021}.

We also evaluated ML schemes \revisionadd{for outlier detection} on a subset of our collected data to assess their capabilities in terms of RBA risk estimation. However, they did not qualify for our use case for the following reasons:
\begin{enumerate*}
    \item Our feature data is categorical with a nominal scale\revisionadd{, i.e., it has no ranking. We did not find a rational reason to rank the categorical feature values in descending order, with each having an individual rank (e.g., is Firefox Mobile a better feature than Firefox Desktop?). Most of these ML schemes, however, }currently only showed good performance with numerical data having an interval scale, e.g., time or sensor measurements~\cite{acien_typenet_2020,rivera_risk-based_2020,deb_actions_2019}. We acknowledge that nominal categorical data \revisionadd{without a ranking} can also be transformed into binary numerical data with one-hot encoding~\cite{scikit_one_2021}. Due to its properties, however, unknown values---such as a new user agent---can only be transformed into one value, i.e., the ``unknown'' category~\cite{burkov_machine_2020}. As this one-hot encoded value has a low scalar distance to the known values, the classification produced inaccurate results in our tests.%
    \item The outlier detection of these mechanisms proved to be suitable only when including the user's own login history in the training~\cite{misbahuddin_design_2017}. Otherwise, as the global feature values in our data were largely distributed, the mechanisms would not detect any outliers in most cases. The data per average user, however, was very sparse at the online service, i.e., four categorical data points per feature (see Section~\ref{sec:data-set}). Based on our tests, this was not enough for sensible outlier classification using \revisionadd{most of} these models, i.e., almost every login risk is high unless all feature values are the same.
    \item In contrast to the selected model, these \revisionadd{ML and non-ML based mechanisms} did not allow for a relation between local and global feature value distributions. This led to inaccurate results with respect to newly established legitimate feature values, e.g., new browser versions across multiple users.
    \item Beyond that, the ML mechanisms need to be trained with new feature values each login\revisionadd{ to optimize the outlier detection for each user. Otherwise, legitimate users would potentially be prompted for re-authentication more often, as new legitimate feature values would be lacking in the trained model}. This, in combination with the classification, was time consuming compared to the Freeman et al. model (e.g., \revisionadd{2.5~s for isolation forest and} 270~ms for OCSVM, compared to 0.003~ms for Freeman et al.). Therefore, we did not consider them practical for large-scale online services in their current form.
\end{enumerate*}

As a result, to focus on a practical and scalable solution for large-scale online services, we chose the RBA model of Freeman et al.~\cite{freeman_who_2016} for further evaluation.
Nevertheless, as its risk score has a numerical interval scale, we still found ML mechanisms useful to optimize the selected RBA model parameters (see Section~\ref{sec:ml-enhanced-rba}). The parameter optimization can be used as a tool for administrators to set up RBA as good as possible in their individual environments.

\section{Freeman et al. RBA Model}\label{sec:rba-model}

The selected RBA model is comparable to a multi-features model derived from observations on the behavior of popular online services~\cite{wiefling_is_2019}. According to the observations, the online services Google, Amazon, and LinkedIn used this model and probably still use it in some form~\cite{wiefling_is_2019}. The model calculates the risk score $S$ for user $u$ and a given feature set $(FV^1, ..., FV^d)$ with $d$ features as:
\begin{equation}\label{eq:rba-without-attack-data}
    S_{u}(FV) = \left( \prod_{k=1}^{d} \frac{p(FV^k)%
    }{p(FV^k | u, legit)%
    } \right) \frac{p(u | attack)}{p(u | legit)}
\end{equation}
The calculation uses the probabilities $p(FV^k)$ of a feature value in the global login history, and $p(FV^k | u, legit)$ of a feature value in the legitimate user's login history. The user login probability is based on the proportion of the user logging in, i.e., $p(u | legit) = \frac{Number\ of\ user\ logins}{Number\ of\ all\ logins}$. Attack data can optionally be included to protect users at risk. As in related work~\cite{wiefling_whats_2021}, we considered all users equally likely being attacked when not having attack data. Thus, we set $p(u | attack) = \frac{1}{|U|}$ with $U$ being the set of users with $u \in U$. We did not use attack data in most of our studies, unless otherwise noted, to be able to compare our results to a related study~\cite{wiefling_whats_2021}. We also assumed that use cases without attack data are more common in practice, especially for medium and small websites that have limited storage and computing capacity.

When considering attack data, we estimated $p(u | attack)$ based on the number of failed login attempts per user%
. We expect that users with a high number of failed login attempts are likely being targeted in credential stuffing or password spraying attacks. If the user is not present in the attack data, we set the probability to the minimum value, i.e., $p(u | attack) = \frac{1}{Total\ number\ of \ logins}$. We did this as it is still possible that other users are attacked.

Furthermore, the equation can be extended to include an attack probability for the used feature values:
\begin{equation}\label{eq:rba-with-attack-data}
    S_{u}(FV) =
    \left( \prod_{k=1}^{d} p(attack | FV^k) \frac{p(FV^k)%
    }{p(FV^k | u, legit)%
    } \right) \frac{p(u | attack)}{p(u | legit)}
\end{equation}
In our use case, $p(attack | FV^k)$ is the number of feature value occurrences in the failed login attempts%
. Similarly, we returned the minimum probability when the feature value was not present in the attack data, i.e., $p(attack | FV^k) = \frac{1}{Total\ number\ of\ attacks}$. Returning a minimum probability is necessary, as we assume that there is always a residual risk of other feature values being used in an attack%
.%

\section{Login Data Set}\label{sec:data-set}

We collected login feature data of 3.3M users collected between February 2020 and February 2021 at a SSO service belonging to \censorchange{the Norwegian telecommunications company Telenor}{a listed telecommunications company\footnote{Company name and business segment omitted during blind review}}. \censorchange{}{The company is one of the world's largest ones in its business segment.} Their customers used the SSO service to access other services provided by the company, e.g., cloud storage or billing information services. The users could authenticate with their username in two ways: Either by entering a password, or by letting the online service's mobile network verify a mobile phone number (see Section~\ref{sec:round-trip-time-feature}). In some cases, users also had to verify their mobile phone number or email address by entering a one-time code (OTP) that was sent to the corresponding number or address.

\subsection{Login Sessions}\label{sec:login-sessions}
\begin{figure}
	\centering
	\includegraphics[width=0.6\linewidth]{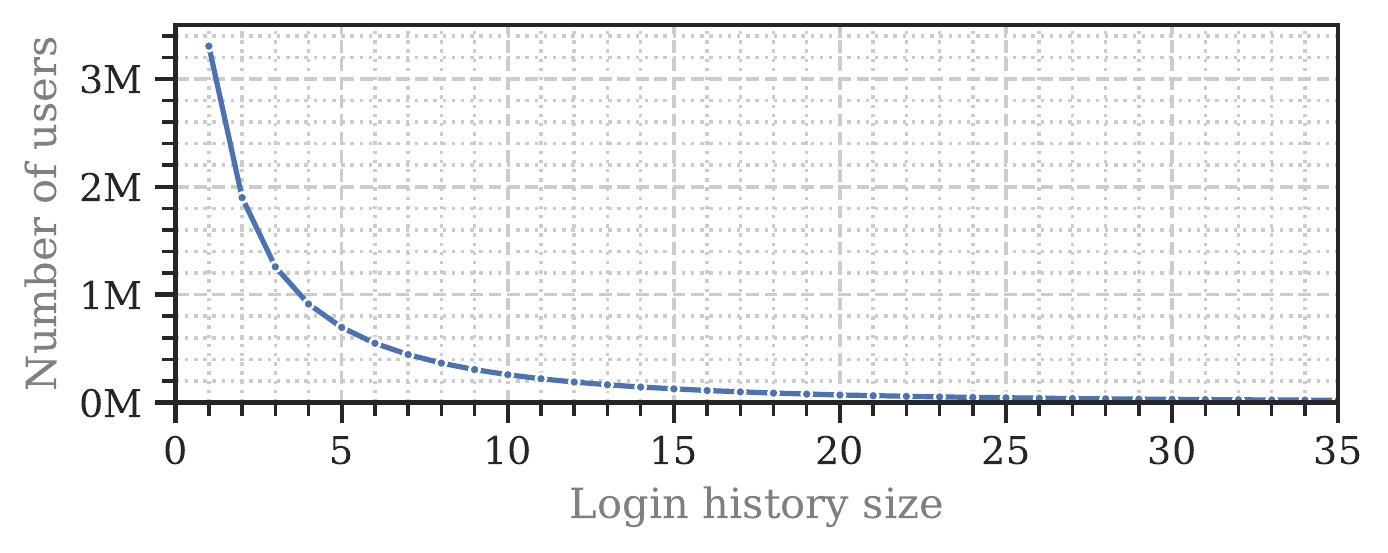}
	\caption{Available login history sizes per user count. We cropped the plot on the x-axis for readability (the maximum size was 5,972).}
	\label{fig:loginhistorysizes}
\end{figure}

We collected feature data of 31.3M login attempts, where 12.5M were successful and 18.8M failed. 25\% of the failed login attempts had a correct username. Out of the failed ones with correct username who proceeded with the login attempt, 90\% failed at the password entry, 9.9\% at the OTP entry, and 0.1\% at the mobile network verification step. Furthermore, 87 successful logins were detected as account takeover by the security incident response team of the company. The users logged in between one and 5,972 times (mean: 3.8, median: 2, SD: 9.35, see Figure~\ref{fig:loginhistorysizes}). The majority of users logged in less than a month (48.3\%) or daily (22.4\%). They used mobile (65.3\%) and desktop devices (34.6\%). The other device types were bots and non-identifiable devices. The operating systems (OS) of desktop devices were Windows (79.2\%), macOS (19.4\%), and Linux (1.4\%) based. Most mobile devices were Android (64.9\%) and iOS (35.1\%). The majority of online browsers were Chrome (59.8\%), Safari (27.4\%), Edge (5.9\%), and Firefox (3.0\%). The order of the browser percentages corresponds to the global and country's browser market share~\cite{w3counter_web_2021}\censorchange{\cite{statcounter_browser_2021}}{\footnote{Citation omitted for blind review}}.

The online service was used by customers residing in \censorchange{Norway}{one country}. We did not collect demographical data for privacy reasons. However, both online service and company target a general audience. Also, the data set's user count is around 2/3 of the country's population. For these reasons, we expect that the data likely represents the country's demographics.

\subsection{Features}
\revisionadd{Due to storage limitations and privacy reasons at the large-scale online service, we had to limit our feature scope for data collection. Previous work identified a subset of useful features out of 247~collected client and server-originated features~{\cite{wiefling_whats_2021}}. We used this work to focus on a subset of the most relevant RBA features that provide good security and usability. As a result, }the data set contained the features \emph{IP address}, \emph{user agent string}, \emph{server-originated RTT}, and \emph{login timestamp}. The online service derived subfeatures from the IP address (ASN \emph{[autonomous system number]}, country, region, and city) and the user agent string (browser/OS name and version, and device type%
). The full plain text IP~address was dropped after this step for privacy reasons. The data set used by the researchers included two privacy-enhanced versions of the IP address, following suggestions from related work~\cite{wiefling_privacy_2021}:
\begin{enumerate*}
    \item The full IP address was salted and hashed by the online service to mitigate potential re-identification. We used this variant in our RBA model. The hashing only transformed the data representation, which is why it did not affect the results produced by the model~\cite{wiefling_privacy_2021}.
    \item The IP address was also included in a truncated form with the last octet removed. We used this version for further analysis to extract a subset of hashed full IP addresses that are potential attack IP addresses. As the company used a very long random salt, the truncated IP still did not allow us to reconstruct the full IP address. However, for data protection and ethical reasons, we did not even attempt this.
\end{enumerate*}

\subsection{Feature Optimization}
Service owners need to optimize the collected features to improve the expected RBA performance. Based on optimizations found in related RBA and browser fingerprinting literature~\cite{wiefling_whats_2021,hurkala_architecture_2014,freeman_who_2016,steinegger_risk-based_2016,alaca_device_2016}, the service extracted %
subfeatures from the IP address and user agent string%
. Administrators also need to weigh the different features to optimize performance. The weighting ensures that certain features are treated with higher priority, e.g., those that are harder to spoof. We calculated the weightings with the method described in Freeman et al.~\cite{freeman_who_2016}. The calculated weightings for IP address (IP address: 0.6, ASN: 0.3, country: 0.1) and user agent (full string: 0.55, browser: 0.29, OS: 0.15, device type: 0.01) were very similar to those of a related study~\cite{wiefling_whats_2021}. Thus, for comparability, we set the same weightings for the IP address (IP address: 0.6, ASN: 0.3, country: 0.1) and user agent (full string: 0.53, browser: 0.27, OS: 0.19, device type: 0.01)\footnote{To assess our results' general validity, we recalculated RQ1's results using the weightings calculated for the data set%
. As these were nearly identical to those using the related study's weightings, this encouraged us to use the latter for a fair comparison.}.

\subsection{Ethical and Legal Considerations}
We do not have a formal IRB process at our university. But besides our ethical considerations below, we made sure to minimize potential harm by complying with the ethics code of \censorchange{the German Sociological Association (DGS)}{a nationwide sociological association} and the standards of good scientific practice of \censorchange{the German Research Foundation (DFG)}{a nationwide research funding organisation\footnote{\label{footnote:orgname-omitted}Organization names omitted during blind review}}. We also made sure to comply with the European General Data Protection Regulation (GDPR)~\cite{european_union_gdpr_2016}.

By using the SSO service, the users agreed in the data collection and evaluation for research purposes. To minimize potential harm, we worked on a privacy preserving data set. All features, and the processes of data pseudonymization and processing were carefully reviewed over more than two months, and approved by the %
legal department of the company. The data was provided by the company only for the research purpose and deleted after research completion. The data was stored on an encrypted external hard drive. Only the researchers had access to the drive and the decryption password. For study reproduction and fostering RBA research, we agreed with the data owner to create a synthesized data set that does not allow re-identification of customers.

\section{Attacker Models}\label{sec:attacker-models}

\begin{figure}[h]
    \centering
    \includegraphics[width=0.65\linewidth]{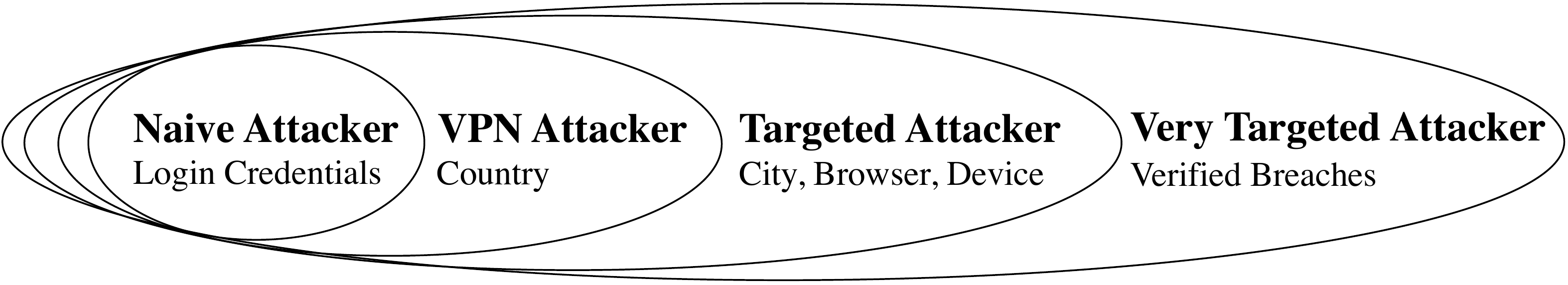}
    \caption{Overview of the attacker models used in the study, with their increasing capabilities to mimic a legitimate user (extended from Wiefling et al.~\cite{wiefling_whats_2021})}
    \label{fig:attacker-models}
\end{figure}

We evaluated the RBA model with four attacker models, with three of them based on known ones in literature~\censorchange{\cite{wiefling_whats_2021,freeman_who_2016,wiefling_evaluation_2020}}{\cite{wiefling_whats_2021,freeman_who_2016}}. Due to verified account takeover data and the relevance in real-world scenarios~\cite{thomas_data_2017,campobasso_impersonation_2020}, we introduce the additional \emph{very targeted attacker}. We describe the attacker models below. %
All attackers possess the login credentials of the victim (see Figure~\ref{fig:attacker-models}).

The \textbf{naive attacker} signs in from a random IP address from somewhere in the world.
We simulated this attacker with IP addresses of failed login attempts that match a data set identifying real-world attacks~\cite{firehol_all_2020}. We expect that a large number of failed login attempts for a user likely indicates that the user was targeted in an attack%
. However, the majority of identified attacks %
came from a particular ASN and a country that was far from the online service's main country. Therefore, detecting those would be trivial. To better diversify the sample, and to simulate global attackers, we only included the most occurring IP address per ASN.
Then, we extracted two types of naive attackers to simulate a wide range of them. Half of the attacks resembled a very naive attacker type comparable to one that randomly tries to enter a login credential. These attacks involved random IP addresses from all 180 countries identified in the data set\footnote{We could not fill in missing countries because we were limited by the pseudonymized IP addresses, i.e., we could only use those that were provided to us.}. The user agents were also randomly selected. The other half of attacks resembled naive attackers that were able to buy additional resources, e.g., servers or botnet. These attacks involved the top IP addresses and user agents having most occurrences in the data set.

The \textbf{VPN attacker} knows the country of the victim. Therefore, the attacker spoofs the IP geolocation using a VPN connection and uses popular user agent strings. The IP address does not necessarily have to belong to a VPN service provider. We simulated this attacker with identified attack IP addresses located in the users' main country~\cite{firehol_all_2020}. As this attacker does not target specific users, we randomly sampled these attack IP addresses from the failed login attempts. The attacker used the most popular user agents from legitimate login attempts%
.

The \textbf{targeted attacker} knows the location, device, and browser of the victim. Therefore, the attacker uses IP addresses and user agent strings that the victim would likely choose. We simulated this attacker with identified attack IP addresses located in the users' main country~\cite{firehol_all_2020}. As the attacker targets specific user groups, we sampled the IP addresses and user agents of the failed login attempts with most occurrences.

The \textbf{very targeted attacker} is the strongest of our attacker models, which only attacks a victim once, but with all information gathered about it. This could, but not necessarily, include accessing the device or the network of the victim, e.g., by malware. We simulated this attacker by replaying the 87 identified successful account takeovers. We did this to check how many of these very targeted attacks would be detected by the RBA system in practice.

\section{Methodology}\label{sec:risk-score-calculations}

We calculated our analysis on a high-performance computing (HPC) cluster with more than 2,400 CPU cores and 7,500~GB RAM. This was crucial, as calculating the risk scores was computational intensive, especially with the large data set used. Computing the results on an eight-core desktop PC would have taken more than a year. The HPC cluster reduced this computation time to approximately 1.5 weeks.

\boldparagraph{Feature Set}
Unless otherwise noted, our RBA model used IP address and user agent string as features. The subfeatures were ASN and country for the IP address, and browser, OS, and device type for the user agent string. We did this to allow a fair comparison to the related study, and since this feature set can be considered the %
state of RBA practice~\cite{wiefling_is_2019,wiefling_whats_2021}. 

\boldparagraph{Optimizations}
The huge data set size largely affected the computation time and limited the computing capacity%
.
Thus, we optimized the RBA algorithm to speed up calculation (see Section~\ref{sec:algorithm-optimization}). The optimizations allowed us to calculate the risk scores for all legitimate logins. However, we still had to limit the attacker calculations to a data subset to be able to obtain the results after a reasonable period of time.

\boldparagraph{Stratification}
Due to the increased storage and computational load caused by the attacker simulation, we calculated the attackers' risk scores for naive, VPN, and targeted attackers on a stratified data set containing 10\% of users. We stratified the user set based on the number of logins to retain the data set properties regarding login histories as good as possible. Stratifying data sets is common practice in ML solutions~\cite{gunopulos_stratification_2011}. To focus our attack detection on real-world attacks, the stratified set included those users who had the highest number of failed login attempts, i.e., likely targeted users.%

To make sure that the stratification did not influence the general validity of our results, we compared the results for RQ1 using all legitimate logins to those using a %
stratified data set. %
The results were identical for the very large majority of login history sizes lower than 100. %
Thus, we considered the stratified sample appropriate for our analysis.

\boldparagraph{Significance Testing}
For statistical tests, we used Kruskal-Wallis (K-W) for the omnibus cases and Dunn's multiple comparison test with Bonferroni correction (Dunn-Bonferroni) for post-hoc analysis. We considered p-values lower than 0.05 as significant.

\section{Evaluating RBA in Practice (RQ1)}\label{sec:testing-rba}

To estimate RBA's performance on a large online service, we repeated a related study~\cite{wiefling_whats_2021} with our data set. The related study estimated RBA characteristics on a small online service. In addition, since the data set was large enough to do this, we further analyzed the number of re-authentication requests depending on the login frequency. We used the same methodology as it was already used for RBA analysis. Beyond that, this allowed us to compare the results.

\subsection{Study Procedure}\label{sec:study-procedure}

We first calibrated the risk scores regarding the percentage of blocked attackers for each attacker model (true positive rate, TPR)~\cite{wiefling_privacy_2021,wiefling_whats_2021,freeman_who_2016}. 
After that, we replayed all legitimate login attempts as they had happened on the online service. For each login attempt, we \begin{enumerate*}
    \item restored the state of the data set at the time of login,
    \item calculated the risk score,
    \item applied the access threshold for the given TPR and attacker model, and
    \item stored the access decision.
\end{enumerate*}

\begin{figure*}[t]
    \centering
    \includegraphics[width=\linewidth]{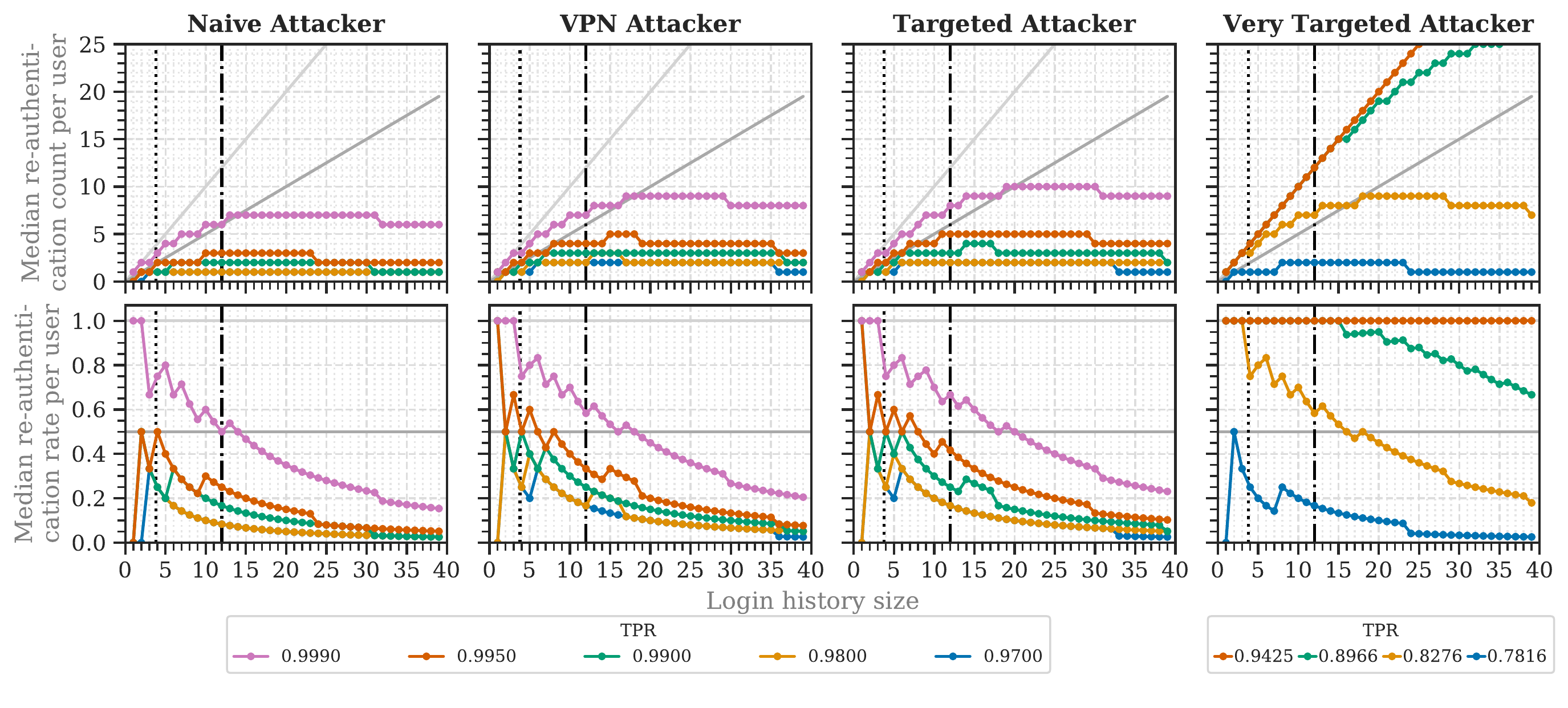}
    \caption{Median re-authentication counts and rates for the four attacker models. The TPR relates to each of the attacker models. For orientation, we added the baseline for 2FA (light grey line), the stable setup threshold (grey line), and the mean login counts for our study (dotted black line) and the study of Wiefling et al.~\cite{wiefling_whats_2021} (dash-dotted black line).}
    \label{fig:average_re_authentication_count_rates_combined}
\end{figure*}

\subsection{Results}

In the following, we present the results ordered by the research questions. A discussion follows after the results.

\boldparagraph{Number of Re-Authentication Requests (RQ1a)}
The users logged in 3.8 times on mean average. Therefore, we considered a user's login history size of four to estimate the re-authentication count for the average user. The median re-authentication count decreased with an increasing login history size (see Figure~\ref{fig:average_re_authentication_count_rates_combined}).

When configuring RBA to block 99.5\% of \textbf{naive attackers}, the average legitimate user was asked for re-authentication every second login. When blocking a lower percentage, users were prompted every fourth login (TPR 0.99-0.95), or not at all (TPR \textless 0.95). When blocking \textbf{VPN} or \textbf{targeted attackers}, legitimate users were prompted every second (TPR 0.995-0.99) or fourth time (0.98-0.90). When blocking at least 82.76\% of \textbf{very targeted attackers}, legitimate users were asked every login attempt.

For a fair comparison to the related study~\cite{wiefling_whats_2021}, we also considered a login history size of 12 to estimate the median login count until re-authentication. In this case, users were asked every fourth login or even less often for TPRs lower than 0.995 (see Table~\ref{tab:logins-until-re-authentication-12}). This is similar to the related study. Summarizing our findings, RBA rarely asks for re-authentication in practice, even when blocking a high number of targeted attackers.

\begin{table}
  \centering
  \caption{Median login count until re-authentication at a login history size of 12}
    \begin{tabular}{@{}lllll@{}ll@{}}
    \toprule
     & \multicolumn{6}{c}{Median logins until re-authentication} \\
    \cmidrule{2-4} \cmidrule{5-7} & Naive & VPN   & Targeted  & & & Very targeted\\
     TPR   & Attacker & Attacker & Attacker & ~~~ & TPR & attacker \\
    \cmidrule{1-4} \cmidrule{6-7}
    0.9990 & 2     & 1.71  & 1.5 & & 0.9425 & 1 \\
    0.9950 & 4     & 3     & 2.4 & & 0.8966 & 1\\
    0.9900 & 6     & 4     & 4 & & 0.8276 & 1.71\\
    0.9800 & 12    & 6     & 6 & & 0.7816 & 6     \\
    0.9500 & 12    & 6     & 6 \\
    0.9000 & $\infty$ & 12    & 12 \\
    \bottomrule
    \end{tabular}%
  \label{tab:logins-until-re-authentication-12}%
\end{table}%

\boldparagraph{Re-Authentication Requests per Login Frequency (RQ1b)}
We classified the users of our data set into five groups having different login frequencies, i.e., daily, several days per week, once a week (weekly), several weeks, and more than 30 days (less-than-monthly). For each user, we calculated the time differences between successive logins. We then classified the user to the group to which the majority of time differences belonged (e.g., the user used the online service mainly on a daily basis). To ensure a sufficient data basis for the classification, we then dropped all users who logged in on less than two different months, two different weeks, and four different days, depending on the class. We based the limit on the median login count and our own observations on the different user groups in the data set. Based on the labels, we calculated the re-authentication counts per login frequency%
.%

\begin{figure}
    \centering
    \includegraphics[width=0.9\linewidth]{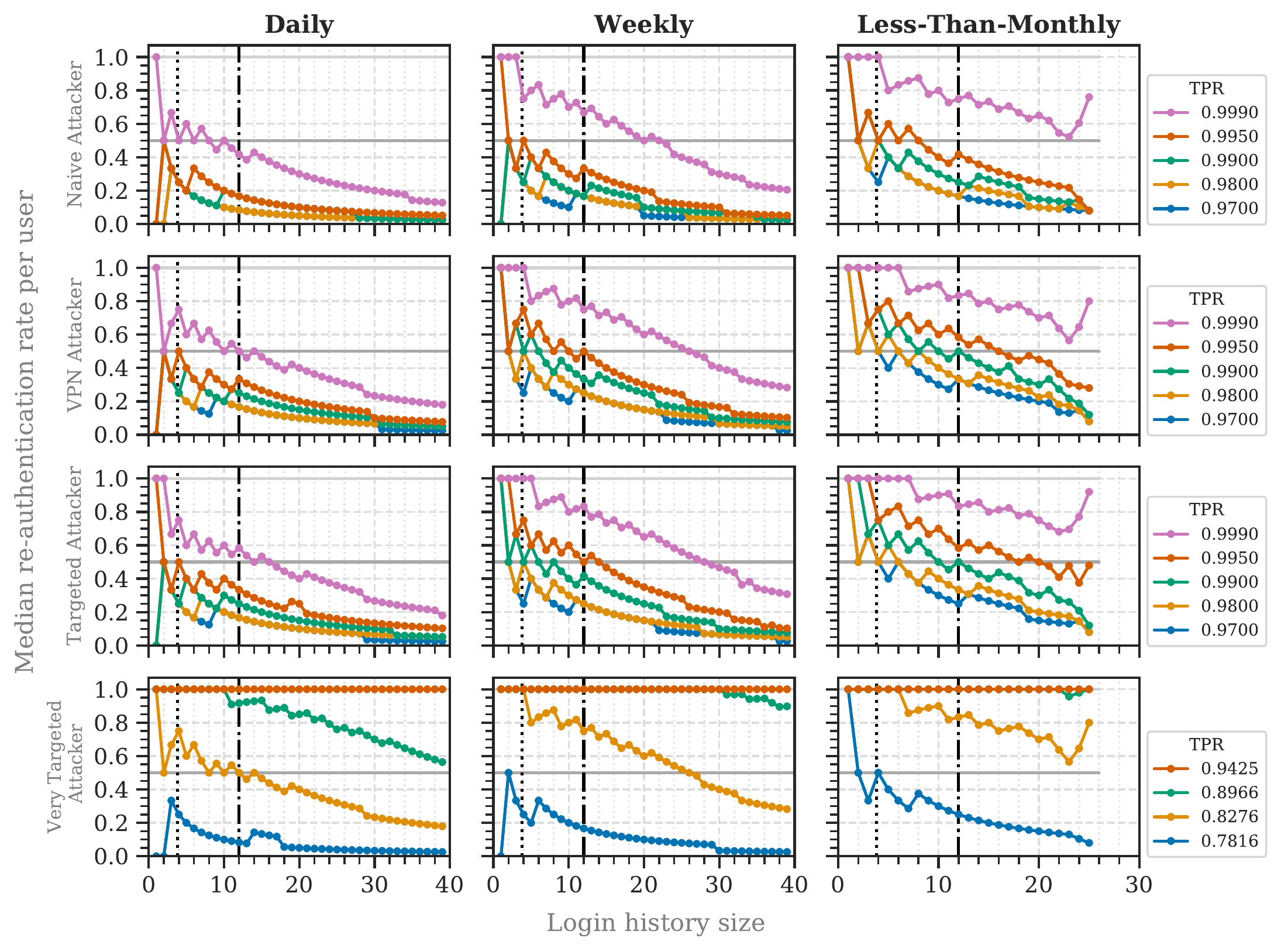}
    \caption{Median re-authentication rates for the attacker models, filtered by login frequency. We added the baseline for 2FA (light grey line), the stable setup threshold (grey line), and the online service's mean login count (dotted black line) for orientation.}
    \label{fig:average_re_authentication_rates_usage_short}
\end{figure}

The median number of re-authentication requests varied significantly between the different login frequency groups (see Table~\ref{tab:login-frequency-significance}). Daily users were asked significantly less for re-authentication than those who logged in several days, and less than several weeks. For instance, when aiming to block 99\% of VPN and targeted attackers, the average daily user had to re-authenticate every fourth login, while weekly or less frequently logging in users had to do it almost every time (see Figure~\ref{fig:average_re_authentication_rates_usage_short}). 
Less-than-monthly users were also prompted for re-authentication significantly more than those who logged in daily, weekly, or several weeks.
Concluding the results, daily users were mostly less asked for re-authentication than those who logged in less frequently.

\begin{table}
    \centering
    \caption{Dunn-Bonferroni p-values for the different login frequency types (TPR: 0.995, targeted attacker). The users logged in daily (\textless 1 day), several days (1-7 days), weekly (7-14 days), several weeks (14-30 days), and less-than-monthly (\textgreater 30 days). We omitted p-values greater than 0.2 for readability. Bold: Significant.}
    \label{tab:login-frequency-significance}
        \begin{tabular}{@{}l@{}rrrrr@{}}
        \toprule
        {} &                      Daily &     Several Days &                     Weekly &    Several Weeks &       \textgreater 30 Days \\
        \midrule
        {Daily               } &                          - &  \textbf{0.0229} &                          - &  \textbf{0.0108} &  \textbf{\textless 0.0001} \\
        {Several Days        } &            \textbf{0.0229} &                - &                          - &                - &                     0.0924 \\
        {Weekly              } &                          - &                - &                          - &           0.1881 &  \textbf{\textless 0.0001} \\
        {Several Weeks       } &            \textbf{0.0108} &                - &                     0.1881 &                - &            \textbf{0.0131} \\
        {\textgreater 30 Days} &  \textbf{\textless 0.0001} &           0.0924 &  \textbf{\textless 0.0001} &  \textbf{0.0131} &                          - \\
        \bottomrule
        \end{tabular}%
\end{table}

\boldparagraph{Required Login History Size (RQ1c)}
In order to achieve RBA's usability gain, users need to notice a difference to 2FA~\cite{wiefling_more_2020}. A baseline for this can be to request re-authentication less than every second time~\cite{wiefling_whats_2021}. Thus, we considered the required login history size as the point after which the median re-authentication rate remained below 0.5.

Following this definition, most TPRs lower than 0.99 required one or no entry in the login history for a stable setup %
(see Figure~\ref{fig:average_re_authentication_count_rates_combined}). The other TPRs required four (naive attacker, TPR 0.995; VPN attacker, TPR 0.99), six (Targeted attacker, TPR 0.99), eight (VPN and targeted attacker, TPR 0.995), and 18 entries (very targeted attacker, TPR 0.8276). This would achieve a baseline setup that, on median, requests re-authentication less than every second time.

However, the required login history size differed based on the login frequency of the users. Even when blocking 99.5\% of all targeted attackers, the median re-authentication rate for daily users remained below the baseline after a login history size of four (see Figure~\ref{fig:average_re_authentication_rates_usage_short}). Thus, in this case, four entries were required for a stable setup targeted at daily users. This was different to the other user types, which required at least eight entries for targeted attackers in this case. The results for the very targeted attacker did not pass the baseline for TPRs greater than 0.89 until a login history size of at least 40.

In general, the login history of each user does not have to contain a large amount of entries for a stable setup. Based on the results, we conclude that storing eight entries is sufficient for a stable RBA setup that blocks 99.5\% of targeted attackers at this online service. When blocking naive or VPN attackers, four to six entries are required in that case.

\begin{figure}
    \centering
    \includegraphics[width=0.8\linewidth]{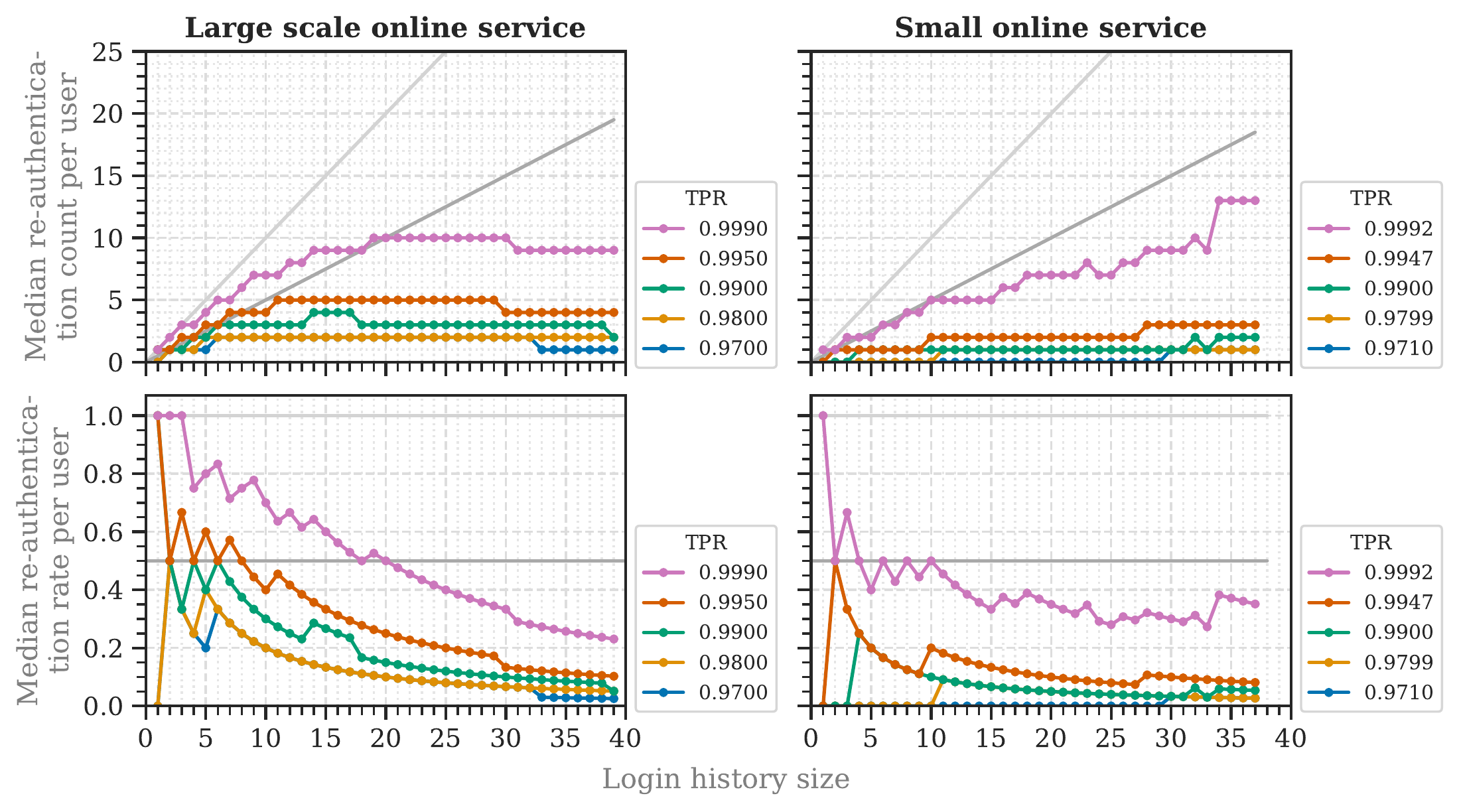}
    \caption{Median re-authentication rates and counts for targeted attackers, compared between a large-scale (our results) and a small online service (Wiefling et al.~\cite{wiefling_whats_2021})}
    \label{fig:average_re_authentication_count_rates_by_online_service}
\end{figure}

\boldparagraph{Comparison to Small Online Service (RQ1d)}
Wiefling et al.~\cite{wiefling_whats_2021} evaluated RBA on a small online service having 780 users and 9,555 legitimate logins. Due to the same setup, we can compare both studies. Our results mainly reflect their findings (see Figure~\ref{fig:average_re_authentication_count_rates_by_online_service}). Users were less requested for re-authentication with an increasing login history size. Although our re-authentication rates mostly declined slower, the main tendency was the same. Therefore, the RBA characteristics did not differ greatly between a large and a small online service.

\subsection{Discussion}

The vast majority of identified attacks on the online service came from a different country than the victim (2.2M incidents, 97\%), representing naive attackers. VPN, targeted, and very targeted attacks using the same country as the victim were less common (58K incidents, 3\%). Therefore, blocking naive attackers already helps to protect against most attacks in practice. Freeman et al.~\cite{freeman_who_2016} reported similar findings regarding botnet attacks, i.e., only 1\% of them had the same country.

The data for very targeted attackers was very sparse%
, which is why the TPRs were very coarse grained. Still, the results show that RBA was able to detect the majority of successful account takeovers (78.16\%) with low re-authentication rates.

The results confirm that RBA can achieve good usability and security after only a few login attempts. However, this varies depending on the user type. Less-than-monthly users had to re-authenticate significantly more than daily users. This is due to the fact that feature values tend to differ more likely after a longer period of time, e.g., because the device or browser was updated, or a new IP address was set~\cite{andriamilanto_guess_2021}. 

Daily to weekly users received less frequent re-authenti\-cation requests, even when RBA was configured for high security. Therefore, RBA can achieve a high acceptance among these users in practice~\cite{wiefling_more_2020}. Less frequent users, however, would be prompted for re-authentication almost every time in this case. The burden of re-authentication is limited here, as users are prompted once per month. Thus, we expect that they tend to accept it~\cite{wiefling_more_2020,crawford_understanding_2014,khan_usability_2015}. However, when being surprised by a re-authentication request after some months, there is a risk of users not being able to solve the re-authentication challenge~\cite{wiefling_more_2020}. This could result in users being annoyed by RBA~\cite{wiefling_more_2020,crawford_understanding_2014,khan_usability_2015}. %
To mitigate this effect, we recommend to set the targeted TPR based on the login frequencies of the general user base. The TPR can be set high for a mostly daily user base, while it has to be lowered for less frequent users. For instance, to achieve a user experience for less-than-monthly users that is comparable to daily users, the TPR needs to be lowered from 0.995 to 0.97 for targeted attackers at this online service (see Figure~\ref{fig:average_re_authentication_rates_usage_short}).

\section{Evaluating RBA With Attack Data (RQ2)}\label{sec:adding-attack-data}

The RBA model offered two possibilities to include attack data (see Section~\ref{sec:rba-model}), which had not been investigated to date. Thus, we evaluated the attack data variations in this Section.
We simulated the login behavior as in the previous study (see Section~\ref{sec:study-procedure}). However, we added the two different attack data variations in our risk score calculation. We outline the procedures together with the results in the following.

\begin{figure}
    \centering
    \includegraphics[width=0.9\linewidth]{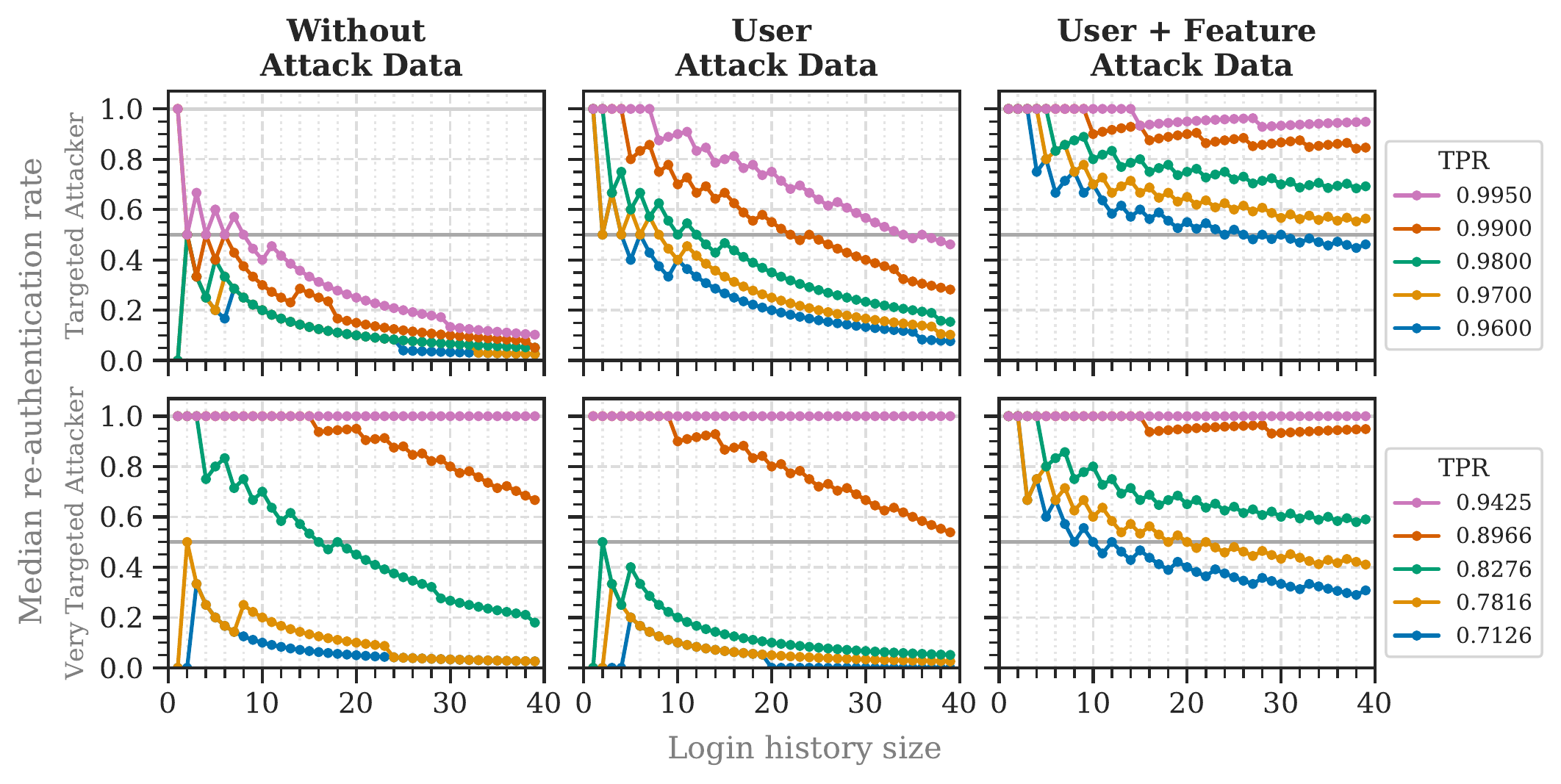}
    \caption{Median re-authentication rates per user for targeted and very targeted attackers when adding user attack data (RQ2a), and both user and feature value attack data (RQ2b).}
    \label{fig:attack_data_re_authentication_rate}
\end{figure}

\boldparagraph{User Attack Data (RQ2a)}
In the first variation, we used %
the user attack probability $p(u | attack)$ calculated from the number of failed login attempts in the attack data. %
Figure~\ref{fig:attack_data_re_authentication_rate} shows the results. When adding the user attack probability, the re-authentication rate significantly increased for naive, VPN, and targeted attackers (p$\ll$0.0001). For very targeted attackers, however, the re-authentication rate significantly decreased for TPRs lower than 0.94 (p$\ll$0.0001). Thus, very targeted attackers could be better distinguished from legitimate users than before, while the opposite was true for the other attackers.

\boldparagraph{User and Feature Value Attack Data (RQ2b)}
In the second variation, we further added probabilities for feature values being used in an attack. Therefore, we used Equation~\ref{eq:rba-with-attack-data} (see Section~\ref{sec:rba-model}) with both the user attack $p(u | attack)$ and the feature attack probability $p(attack | FV^k)$ calculated from the failed login attempt data. %

When also including the feature attack probability, the re-authentication rates significantly increased for all attacker models (p$\ll$0.0001, see Figure~\ref{fig:attack_data_re_authentication_rate}). Thus, attackers could not be distinguished from legitimate users as good as before.

\boldparagraph{Discussion}
Using the user attack data in the RBA model reduced re-authenti\-cation requests when blocking \emph{very targeted attackers}. As the targeted users were likely attacked before, the risk scores increased for them. However, the risk scores decreased for attacks on those users who were not attacked before. Thus, the usability performed worse for the other attacker models that targeted a wide range of users. Therefore, the decision to include attack data should be made with great caution. Related work suggests that other metrics like social media followers~\cite{freeman_who_2016} or frequently guessed passwords~\cite{tian_stopguessing_2020} could help to identify %
accounts that should receive higher protection. This, however, requires further research.

Adding feature attack data made it harder for the RBA system to distinguish between legitimate users and attackers. Attackers identified in the data set likely chose popular feature values, e.g., the user agent. Thus, legitimate users having these feature values received a high risk score. Attackers trying unpopular feature values, however, likely received risk scores in the range of legitimate users using popular feature values. As a result, legitimate users could not be distinguished from attackers as good as before. Considering this, we do not recommend using the tested feature attack data variant in the RBA model to calculate the risk score.

\section{Login History Minimization (RQ3)}\label{sec:login-history-minimization}

Over the last few years, more and more data protection regulations aimed to protect users from massive data collection by online services. These regulations, like the CCPA~\cite{california_ccpa_2018} and GDPR~\cite{european_union_gdpr_2016} suggest to limit the data storage as good as possible. In terms of RBA, this means that the feature data has to be disposed as soon as it does not serve the purpose of RBA any more~\cite{wiefling_privacy_2021}. One measure to minimize data is to delete global login history entries after a fixed amount of months~\cite{wiefling_privacy_2021}. Such a data limiting procedure is also useful to maintain an acceptable authentication speed~\cite{wiefling_whats_2021}. To estimate the potential of the minimization approach, we evaluated it below.

\subsection{Study Procedure}
We simulated the login behavior as in the first study (see Section~\ref{sec:study-procedure}). For each login, however, we only kept the global login history within three different monthly ranges (1, 3, and 6 months). We selected these ranges to estimate the scope of possible effects when minimizing the login history. We further included the full data set results (12 months) for comparison. To enable RBA protection for all users, including less-than-monthly ones, we kept at least one entry in the user's login history.

We attacked each user at the point at which the user last logged in, to observe the possible RBA protection with login history minimization. We had to select a specific point in time for the attacks due to the limited computing and storing capacity. Therefore, we selected the point with the highest login history size distribution, to observe a wide range of effects. Otherwise, when selecting later points in time, the login history sizes of all users would approach one. Thus, we would not likely be able to observe daily to several weeks users in the results.

\subsection{Results}

\begin{figure}
    \centering
    \includegraphics[width=\linewidth]{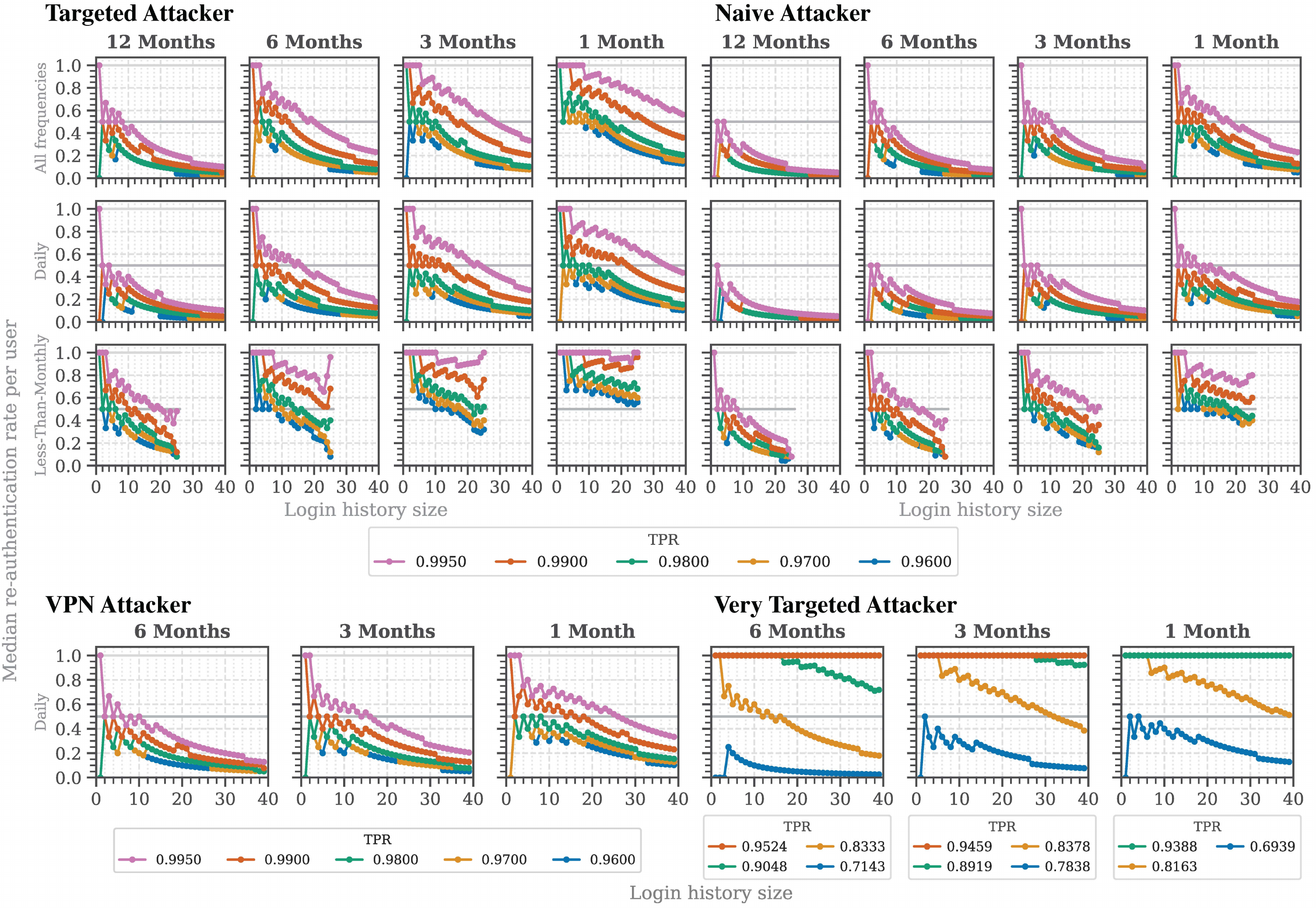}

    \caption{RQ3: Median re-authentication rates by login frequencies and attacker model when minimizing the login history.}
    \label{fig:average_re_authentication_rates_minimization}
    
    \label{fig:average_re_authentication_rates_minimization-targeted}
    \label{fig:average_re_authentication_rates_minimization-naive}
    \label{fig:average_re_authentication_rates_minimization_daily}

\end{figure}

For all attacker models, the median re-authentication rate per user increased when minimizing the global login history (see Figure~\ref{fig:average_re_authentication_rates_minimization}). The differences between the monthly ranges were significant for all attackers (p$\ll$0.0001), except for the naive attacker between three and both six and twelve months (TPR 0.995; see Figure~\ref{fig:average_re_authentication_rates_minimization-naive}). When keeping less than six months, the median re-authentication rate for daily users increased after a few logins (VPN: 6 months, very targeted/targeted: $\leq$3 months, naive: 1 month).

\subsection{Discussion}

The results show that, in terms of naive attackers, reducing the global login history to three months can be possible without significantly influencing the median re-authentication count. However, this is different when blocking more intelligent attackers. For targeted attackers and a high TPR ($\ge$0.99), daily users were prompted for re-authentication almost every login when keeping less than six months. Users would not accept such frequent re-authentication~\cite{wiefling_more_2020,crawford_understanding_2014,khan_usability_2015}. Therefore, when aiming to block a high number of attackers, keeping between 6 and 12 months of login history seems reasonable for a stable setup on the online service.

Our results showed the login history minimization's potential to increase user privacy while maintaining usability and security. Further research should investigate the effects of limiting both the user's login history and the global login history to a maximum size. These approaches could also potentially reduce the authentication time and required storage.

\section{ML-Based RBA Parameter Optimization (RQ4)}\label{sec:ml-enhanced-rba}

In order to optimize the usability and security properties of RBA, administrators need to analyze the risk scores of legitimate users and attackers to set a suitable access threshold. This is, however, a time-consuming and complex process. For this reason, we analyzed whether the RBA configuration process can be automated and improved with ML mechanisms to achieve a good RBA setup in short time.

\subsection{Dynamic Access Threshold}

Current RBA literature considered the access threshold as a static, one-value component to distinguish between two risk categories~\cite{freeman_who_2016,steinegger_risk-based_2016,rivera_risk-based_2020,wiefling_whats_2021}. However, when analyzing the risk scores for legitimate users and attackers, we discovered that they declined with an increasing login history size%
. Therefore, for optimized usability and security, the access threshold can be set higher at the beginning, and follow this decline with an increasing login history size. Based on our findings, we introduce a dynamic access threshold, which uses different access thresholds based on the user's login history size. To select appropriate threshold values, a ML-based approach can assist administrators in doing so.

\boldparagraph{Overview}
\begin{figure}
    \centering
    \includegraphics[width=0.7\linewidth]{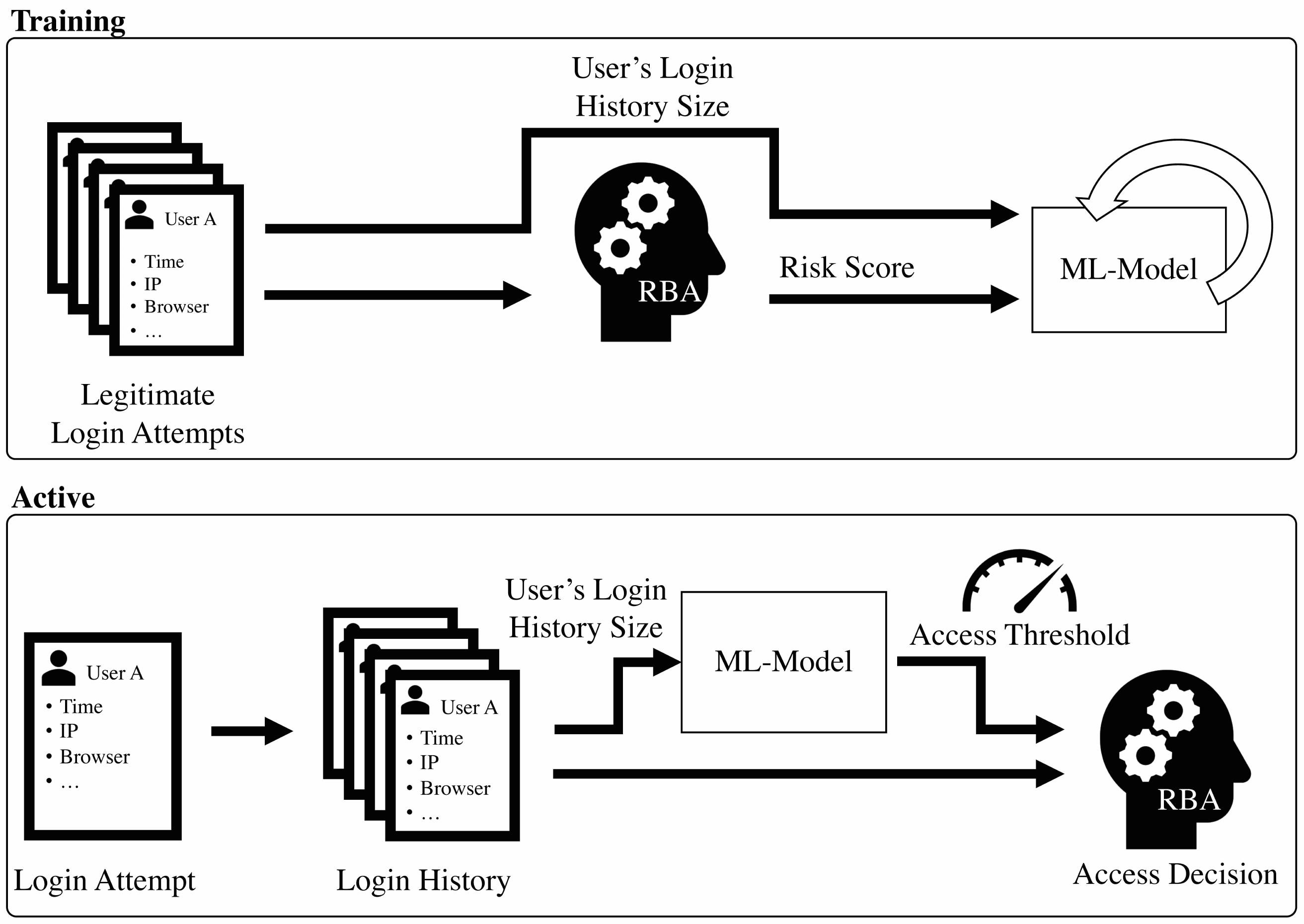}
    \caption{Workflow of the ML-based RBA parameter optimization using the dynamic access threshold}
    \label{fig:dynamic_access_threshold}
    \label{fig:dynamic_access_threshold_learning}
\end{figure}
Our ML-based approach works as follows (see Figure~\ref{fig:dynamic_access_threshold}). We first calculate the risk scores and the user's corresponding login history size for a number of legitimate login attempts. After that, a ML model learns the risk scores based on the login history size. As a result, the ML model can generate a risk score that is closest to most legitimate users, i.e., an optimized access threshold. Assuming that legitimate users have lower risk scores than attackers, %
this can %
reduce the re-authentication count while not decreasing security to a large extent.

After training the ML model, the RBA algorithm extracts the ML-generated dynamic access threshold based on the login history of the user and uses it for the access decision.

\boldparagraph{Access Threshold Calculation}
The risk scores for legitimate users declined fast for the first few login history sizes and then followed a near-linear decline%
. To consider this progression over the login history size, we trained the dynamic access thresholds with regression models. We created and tested three methods to calculate the dynamic access thresholds. These were
\begin{enumerate*}
    \item[(i, Linear)] a linear regression model to follow a linear progression,
    \item[(ii, Poly)] a polynomial regression model to follow a fast decline, and
    \item[(iii, Hybrid)] a hybrid of both models to follow the expected risk score development over the login history size.
\end{enumerate*}

The risk scores might contain outliers that increase the generated access thresholds, and thus decrease the security properties. To counteract this, we cleaned the data from the highest 5\% of scores. We based this threshold on the observed risk scores. RBA risk scores generated by our tested model cannot go lower than zero. To make sure that the generated dynamic access thresholds were valid, we only used those that were greater than zero. If this was not the case, we kept the access threshold of the previous login history size. To achieve the best possible security and adoption to the expected risk scores, the hybrid model took the lowest access threshold of both models.

\subsection{Studies}

We designed two studies to estimate the performance of the ML-enhanced RBA model in practice. We trained the model on sample slices of 100K consecutive legitimate logins, which were less than 1\% of our data set. As the login history size ranged between one and 5,972, we considered 100K as a reasonable size to include multiple samples of all sizes as good as possible. We acknowledge that random sampling a training set from the complete data set is often used in ML applications~\cite{james_introduction_2013}. However, we assume that using consecutive logins is more realistic for RBA in practice, as we can only sample logins that happened in the past. Also, the available historical login data is often sparse in practice, so random sampling is likely not possible.

\boldparagraph{Study 1: TPR Stability}
We first estimated the range of TPRs that each of the three ML models achieved in practice. We trained the models with multiple sample slices and calculated the TPRs for each of the attackers. We extracted the 100K sample slices in consecutive 500K steps from the data set, to approximate the performance over the whole data set.

After training, the Hybrid model achieved higher TPRs than the Linear and Poly model (see Table~\ref{tab:dynamic-access-thresholds-tprs} in Appendix~\ref{appendix:ml-results}). The TPRs of all models varied between 0.93 and 0.997 for naive attackers, 0.86 and 0.996 for VPN and targeted attackers, and 0.65 and 0.87 for very targeted attackers. %
Therefore, the Hybrid model achieved a stronger security than the other two models.

\boldparagraph{Study 2: Re-Authentication Rates}
To identify how often the dynamic access threshold requests for re-authentication in a practical example, we trained the ML models on the first 100K legitimate logins, and calculated the re-authentication counts and rates. We selected the first logins to estimate the performance over the course of one year.

\captionsetup[subfigure]{position=top}
\begin{figure}[t]
    \centering
    \subfloat{
	    \includegraphics[width=0.62\linewidth]{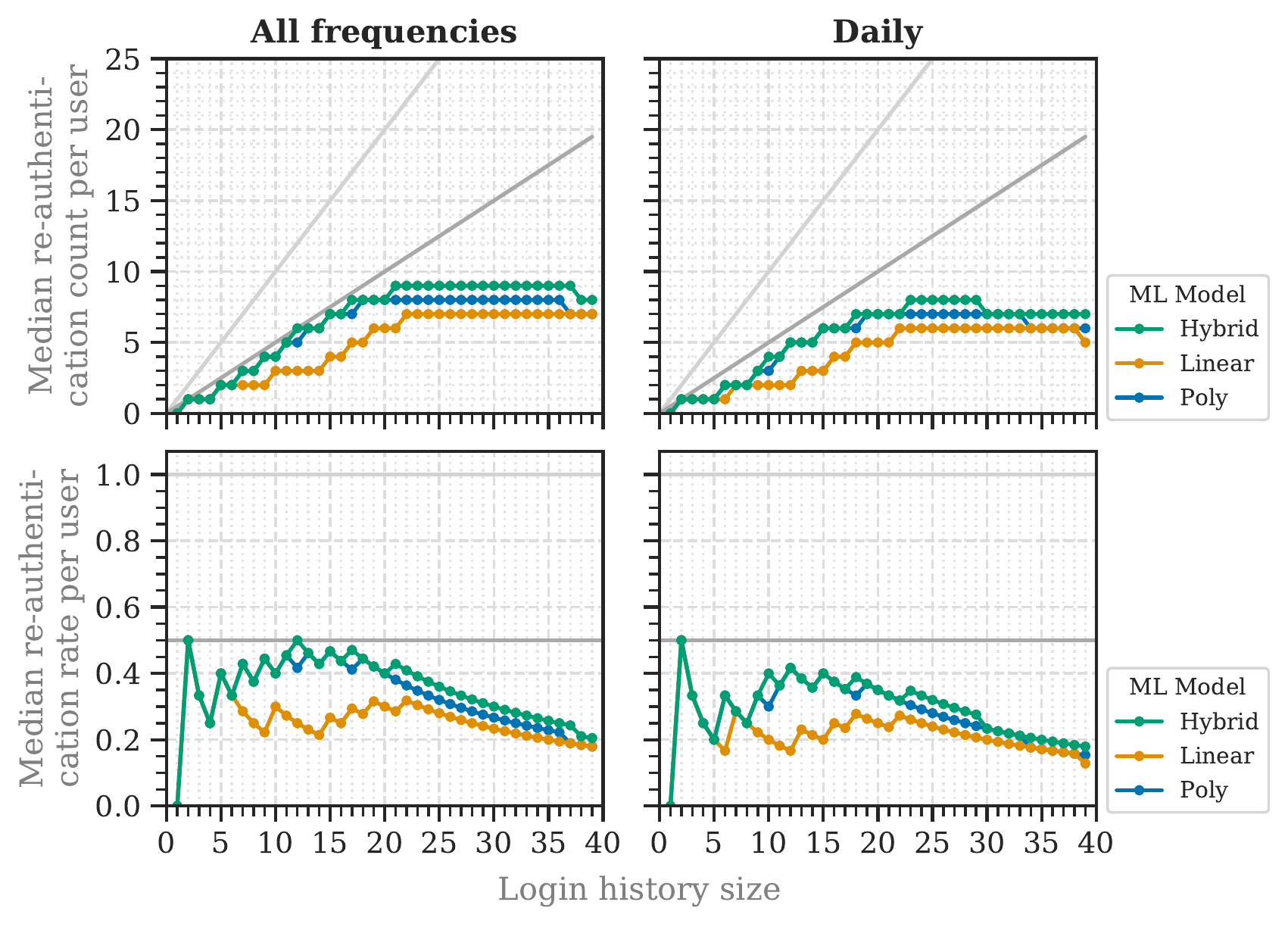}
	}
    \hfill
    \subfloat{
	    \resizebox{0.35\linewidth}{!}{
	    \begin{tabular}{@{}llll@{}}
    	    & \\
	        \toprule
	         & \multicolumn{3}{c}{TPR} \\
	        \cmidrule{2-4} Attacker Model &  Linear &   Poly &  Hybrid \\
	        \midrule
	        Very Targeted Attacker &  0.8276 & 0.8391 &  0.8391 \\
	        Targeted Attacker      &  0.9936 & 0.9954 &  0.9956 \\
	        VPN Attacker           &  0.9927 & 0.9953 &  0.9956 \\
	        Naive Attacker         &  0.9961 & 0.9966 &  0.9967 \\
	        \bottomrule
	    \end{tabular}}
	}

    \caption{RQ4, Study 2: Median re-authentication counts and rates by login frequency (left) and achieved TPRs (right) for the dynamic access thresholds generated by three different ML models. The ML models were trained on the first 100K legitimate logins. The differences between all three ML models were significant.}
    \label{fig:average_re_authentication_count_rates_ml}
\end{figure}
\captionsetup[subfigure]{position=bottom}

In contrast to the static access threshold, the dynamic access threshold never exceeded a 0.5 median re-authentication rate per user at TPRs higher than 0.995 (see Figure~\ref{fig:average_re_authentication_count_rates_ml}). %
Therefore, the dynamic access threshold reduced re-authentication requests compared to the static threshold. However, the median re-authentication rate varied significantly between the different ML-models generating the dynamic access threshold.
The Linear model requested significantly less re-authentication than the Poly and Hybrid model (p$\ll$0.0001). The Hybrid model requested significantly more re-authentication than the Linear and Poly method (p$\ll$0.0001).

\subsection{Discussion}

The dynamic access threshold performed better than the static threshold, as it achieved lower re-authentication rates at the same TPRs%
. %
For a high usability, we recommend using the linear model%
, as its TPRs were similar to those of the other models but with significantly lower re-authentication rates.

The results show that the ML-based parameter optimization can serve administrators as a basic setup to balance RBA's usability and security in little time. However, the achieved TPR varied based on the trained input, which is probably not desired by online service operators. This is why we consider the generated dynamic access thresholds more as a starting point for administrators. They can take these thresholds and adjust them to their needs, e.g., reducing them to increase security and re-authentication count. They can also adjust the training set to drop fewer high risk score outliers to increase the generated access thresholds in general.

\section{Performance (RQ5)}\label{sec:performance}

Using RBA with large global login history sizes can greatly increase the risk score calculation time~\cite{wiefling_whats_2021}. To be able to compute the results %
in reasonable time, we optimized the RBA algorithm. We describe our optimizations including an analysis in the following. The optimizations are also applicable to productive environments.

\subsection{Algorithm Optimization} \label{sec:algorithm-optimization}

To calculate the probabilities in the risk score (see Section~\ref{sec:rba-model}), the RBA model needs to execute multiple database queries to count the number of feature occurrences%
. The time for each query increases with the global login history size~\cite{wiefling_whats_2021}. Therefore, calculating the risk score for one login attempt took around 5.6~s when using all 12.5M login history entries. Since such a delay is not feasible in a real-world scenario~\cite{borzemski_impact_2018}, we optimized the %
algorithm with hash tables~\cite{liu_empirical_2014}, one for each feature. The hash table stored the number of occurrences per feature value as key value pairs, i.e., \emph{feature value} $\rightarrow$ \emph{number of occurrences}. After each successful login attempt, the hash table values were updated by increasing the feature value counts of those features belonging to the login attempt. Since this replaced the computational intensive query with a simple addition, i.e., $current\ value + 1$, it reduced the computation time to a large degree. Calculating the risk score using hash tables took around 0.2~s, which was a 28x speedup.

We verified through partial calculations that the optimized risk score calculations were identical to the unoptimized version. We calculated both risk scores types on multiple slices of 5K login attempts at random positions of the data set and compared the results.

\subsection{Analysis}

\begin{figure}[t]
    \centering
    \includegraphics[width=0.75\linewidth]{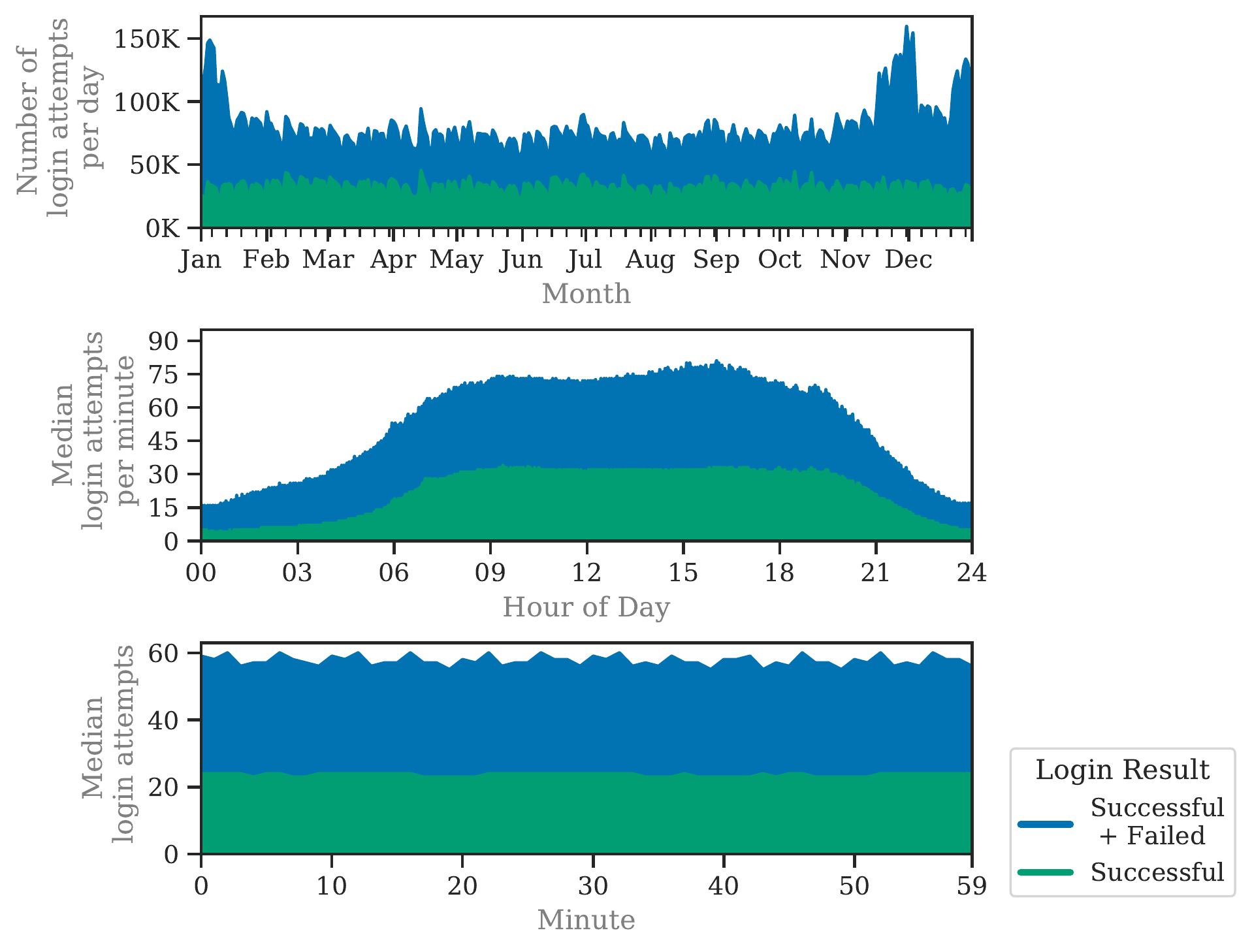}
    \caption{Login attempts on the online service over the year (top), and on a daily (middle) and hourly basis (bottom). The service processed a median of 74.8K login attempts per day (SD: 18.6K), and 58 logins per minute (SD: 28).}
    \label{fig:login_attempts_per_day_and_minute}
\end{figure}

The online service processed a median of more than 74K login attempts per day, with peaks over 150K around Christmas and New Year (see Figure~\ref{fig:login_attempts_per_day_and_minute}). The peaks resulted from an increase in failed login attempts, also from identified attack IP addresses. 
Such high numbers may influence the RBA authentication time per user, especially with multiple login attempts in parallel.%

To get an estimate of the expected authentication time per login attempt, we measured the risk score calculation time based on the global login history size%
~\cite{wiefling_whats_2021}. Due to a large calculation time, measuring both unoptimized and optimized RBA algorithms for all 12.5M login history sizes multiple times was not feasible. To consider the increasing calculation times, we measured the risk score calculation multiple times with global login histories that increased in 500K steps, and calculated the linear regression line. As the authentication time increases linearly with the global login history size~\cite{wiefling_whats_2021}, a linear regression was feasible for our analysis. We measured on a server with Intel Xeon Gold 6130 processor (2.1 GHz, 64 cores), 480 GB SSD storage, and 64 GB RAM. We determined the effect sizes based on Cohen~\cite{cohen_statistical_1988}.

Without optimization, the regression yielded to $y=58.7327 + 0.00044 \cdot x$ with a large effect size ($R^2$=0.95; f=4.40; p$\ll$0.0001), where $y$ is the time in ms and $x$ the global login history size. With hash table optimization, the regression line resulted in $y=69.1459 + 0.00002 \cdot x$ and a large effect size ($R^2$=0.99; f=9.25; p$\ll$0.0001).

\begin{figure}[t]
    \centering
    \includegraphics[width=0.7\linewidth]{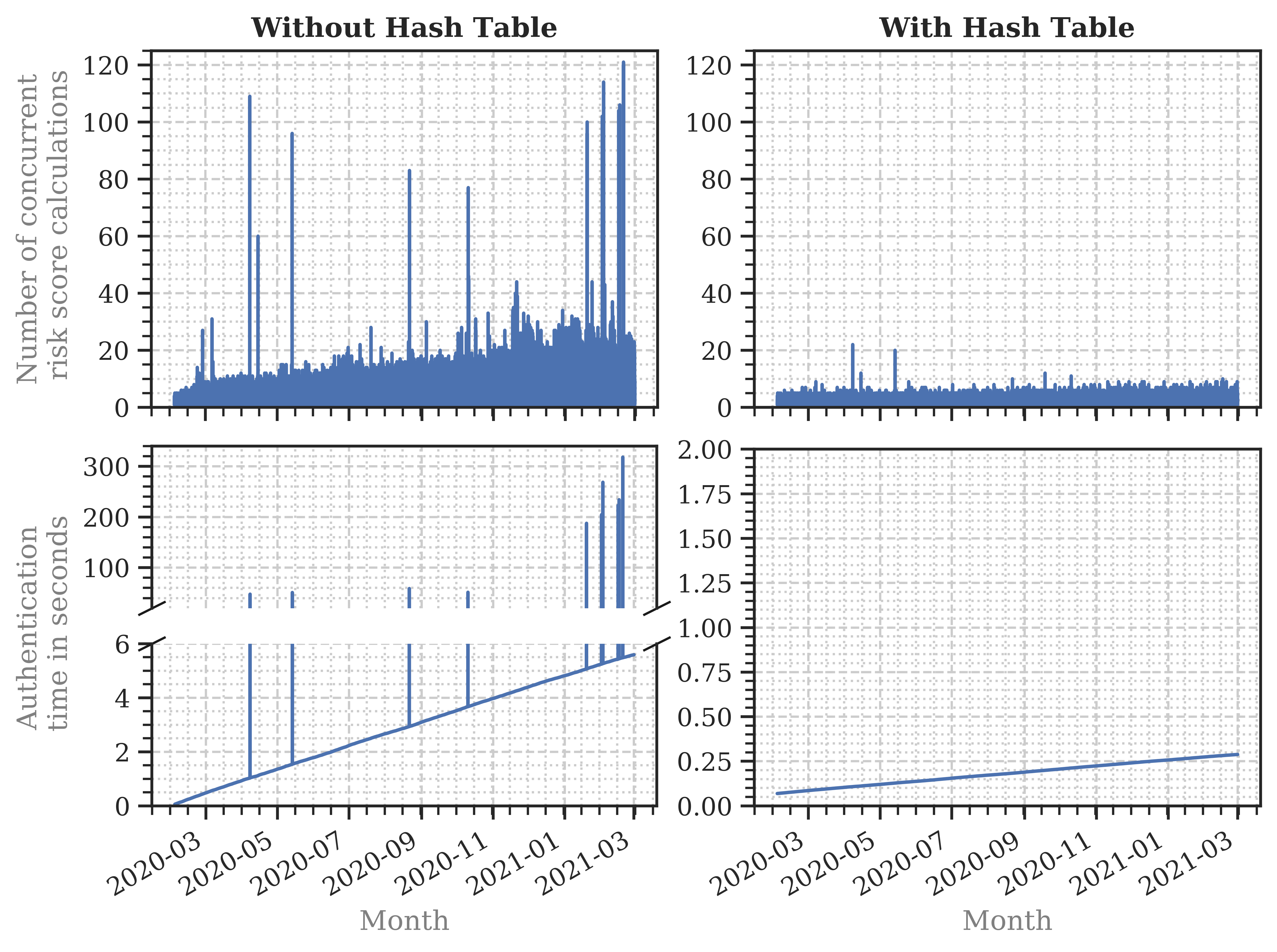}
    \caption{Concurrent risk score calculations and expected authentication times within the observation period at the online service. We calculated the times for a 64 core processor using the regression lines representing the average authentication times per global login history size. Using a hash table significantly decreased the authentication time.}
    \label{fig:risk_score_calculation_times}
\end{figure}

\boldparagraph{Authentication Time in Practice}
We used the resulting regression lines and the login timestamps to calculate the estimated risk score calculation time ranges, i.e., start and end of calculation. Based on their intersections, we identified the number of risk scores that were calculated at the same time (see Figure \ref{fig:risk_score_calculation_times}). The number of concurrent risk score calculations reduced significantly when using the hash table (p$\ll$0.0001).

We assume that one processor core can calculate one risk score at at time, so the authentication time might slow down with multiple login attempts at similar times. To estimate a realistic authentication delay for our data set, we calculated the authentication time for a server having 64 processor cores. In this case, the median authentication time was 193~ms (SD: 64~ms) with a hash table, and 3.2~s (SD: 2.77~s) without a hash table (see Figure~\ref{fig:risk_score_calculation_times}).

\boldparagraph{Required Storage}
The memory storage for our hash tables depends on the number of unique feature values in the data set. In practice, users had an intersection of some feature values, so the required storage was much smaller than the full data set. Also, the feature values were hashed in the table, so they required much less memory than the raw string values. As a result, the required memory for the hash table was very small, even after all recorded logins were added %
(see Table~\ref{tab:hashtable-storage}).

\begin{table}[t]
    \centering
    \caption{Memory usage for the hash table with all 12.5M legitimate logins. The hash table required 1\% of the login history memory (8.28 GB).}
    \begin{tabular}{@{}lr@{\hskip 4em}lr@{}}
    \toprule
             Feature & Storage &    Feature & Storage \\
    \midrule
              User ID & 50.42 MB &  IP Address & 35.08 MB \\
                  ASN &  0.12 MB &     Country &  0.003 MB \\
    User Agent String &  3.89 MB &     Browser &  0.05 MB \\
                   OS &  0.01 MB & Device Type &  0.0001 MB \\
    \midrule
    Total & 89.56 MB &             &          \\
    \bottomrule
    \end{tabular}
    \label{tab:hashtable-storage}
\end{table}

\subsection{Discussion}
Not using a hash table greatly increased the authentication time. It increased even more with a large amount of login attempts. In the extreme case, when the concurrent login attempts exceeded the processor core count, the authentication times were similar to a denial of service attack. This also shows that unoptimized RBA can be vulnerable to such attacks with little effort. To significantly improve the authentication performance on large-scale online services, we therefore recommend using the hash table. %
Due to the low memory footprint, the hash table can be stored in RAM for fast performance (mean: 0.003 ms with hash table, 35 ms without hash table)\footnote{Time to get a userid's entry count on the full data set, i.e., the largest delay possible in our use case scenario\label{footnote:performance}}. Using a relational database, e.g., MariaDB, for storage would slow down the overall query time, but the hash table would still increase the %
performance (mean: 3.5 ms with, 8.9 s without)\footnotemark[\getrefnumber{footnote:performance}].

\section{Round-Trip Time Feature (RQ6)}\label{sec:round-trip-time-feature}

Wiefling et al.~\cite{wiefling_whats_2021} proposed a server-originated round-trip time (RTT) feature based on the WebSocket technology~\cite{melnikov_websocket_2011} to estimate device location. The server requests a data packet (ping frame) from the client and measures the time until the response (pong frame). Only devices near the server location can achieve low RTTs\revisionadd{, which is why this is hard to spoof for attackers without high effort, i.e., they need access to a device physically located in the target area.} Thus, this feature can help to determine whether the client's device location is really in the indicated region, or spoofed by VPNs or proxies~\cite{campobasso_impersonation_2020,abdou_secure_2018}. \revisionadd{Content Delivery Networks (CDNs) can even improve the reliability of RTT, as edge nodes close to the client's device are also considered for the measurement~{\cite{wiefling_whats_2021}}. Concurrent and independent work confirmed this potential}~\cite{kohls_verloc_2022}. For most re-identification attacks with leaked data, the RTT is useless, because it depends on the server location, which is widely distributed in practice. Therefore, the RTT also has potential to be a privacy-enhancing alternative to the sensitive IP address feature~\cite{wiefling_privacy_2021}. To investigate its potential as a location verifier and as a drop-in replacement for the IP address feature, we evaluated the RTT feature's abilities.

The online service did not collect RTTs for all users during login. However, the online service offered mobile users the option to verify their mobile phone number when connected to the service's network provider. In this case, the online service sent a request to the mobile phone, to which the phone responded to. The service measured the time (in ms) for the response, which can be considered a RTT variant. The RTT was recorded in 5.1M login attempts (2.5M successful, 2.6M failed), which were 17.6\% of all login attempts.

\subsection{RTT as a Location Verifier}

We %
investigated the RTT's potential to verify locations that were given by the IP-based geolocation. In contrast to Wiefling et al.~\cite{wiefling_whats_2021}, the RTT was not measured five times during each login, so the RTTs may largely vary due to mobile connectivity. To mitigate this effect, we only considered cities with at least ten measurements. We describe our studies to measure the RTT's ability to determine countries and regions below. 

\begin{figure}[t]
        \includegraphics[width=0.65\linewidth]{images/statistics/rtts_by_country\censorchange{}{-anonymized}}
    \caption{Median RTTs by country. The RTTs were measured on mobile devices. \censorchange{}{Some values were changed for blind review.}}
    \label{fig:rtts_by_country}
\end{figure}

\begin{table}[t]
	\centering
	\caption[Dunn-Bonferroni p-values for the RTTs]{Dunn-Bonferroni p-values for the RTTs. We omitted p-values greater than 0.2 for readability. Bold: Significant.}
	\subfloat[Countries]{%
		\resizebox{0.6\linewidth}{!}{%
			\begin{tabular}{@{}lllllllll@{}}
				\toprule
				{} &                         BD &      DE &               MM &                         MY &                         \censorchange{NO}{DK} &                         PK &               SE &                         US \\
				\midrule
				BD &                          - &       - &                - &  \textbf{\textless 0.0001} &  \textbf{\textless 0.0001} &            \textbf{0.0074} &           0.1686 &                          - \\
				DE &                          - &       - &                - &                     0.1381 &                          - &                          - &                - &                          - \\
				MM &                          - &       - &                - &            \textbf{0.0038} &                          - &                          - &           0.1082 &                          - \\
				MY &  \textbf{\textless 0.0001} &  0.1381 &  \textbf{0.0038} &                          - &            \textbf{0.0410} &  \textbf{\textless 0.0001} &                - &  \textbf{\textless 0.0001} \\
				\censorchange{NO}{DK} &  \textbf{\textless 0.0001} &       - &                - &            \textbf{0.0410} &                          - &  \textbf{\textless 0.0001} &                - &  \textbf{\textless 0.0001} \\
				PK &            \textbf{0.0074} &       - &                - &  \textbf{\textless 0.0001} &  \textbf{\textless 0.0001} &                          - &  \textbf{0.0002} &                          - \\
				SE &                     0.1686 &       - &           0.1082 &                          - &                          - &            \textbf{0.0002} &                - &            \textbf{0.0005} \\
				US &                          - &       - &                - &  \textbf{\textless 0.0001} &  \textbf{\textless 0.0001} &                          - &  \textbf{0.0005} &                          - \\
				\bottomrule
				\censorchange{}{\multicolumn{9}{@{}l}{\rule{0pt}{4ex}\small \emph{Country names were changed for blind review.}}}
			\end{tabular}%
			\label{tab:significance-rtt-country}
		}%
	}
	\quad
	\subfloat[Counties of main country]{%
		\resizebox{\linewidth}{!}{%
			\begin{tabular}{@{}llllllll@{}}
				\toprule
				{} & \censorchange{Innlandet}{County 1} & \censorchange{Nordland}{County 2} & \censorchange{Troms og Finnmark}{County 3} & \censorchange{Trøndelag}{County 4} & \censorchange{Vestfold og Telemark}{County 5} & \censorchange{Vestland}{County 6} & \censorchange{Viken}{County 7} \\
				\midrule
				\censorchange{Innlandet}{County 1}            &                                  - &                   \textbf{0.0023} &                            \textbf{0.0052} &                    \textbf{0.0196} &                                             - &                            0.1161 &                              - \\
				\censorchange{Nordland}{County 2}             &                    \textbf{0.0023} &                                 - &                                          - &                                  - &                               \textbf{0.0334} &                                 - &      \textbf{\textless 0.0001} \\
				\censorchange{Troms og Finnmark}{County 3}    &                    \textbf{0.0052} &                                 - &                                          - &                                  - &                               \textbf{0.0474} &                                 - &                \textbf{0.0003} \\
				\censorchange{Trøndelag}{County 4}            &                    \textbf{0.0196} &                                 - &                                          - &                                  - &                                             - &                                 - &                \textbf{0.0007} \\
				\censorchange{Vestfold og Telemark}{County 5} &                                  - &                   \textbf{0.0334} &                            \textbf{0.0474} &                                  - &                                             - &                                 - &                              - \\
				\censorchange{Vestland}{County 6}             &                             0.1161 &                                 - &                                          - &                                  - &                                             - &                                 - &                \textbf{0.0024} \\
				\censorchange{Viken}{County 7}                &                                  - &         \textbf{\textless 0.0001} &                            \textbf{0.0003} &                    \textbf{0.0007} &                                             - &                   \textbf{0.0024} &                              - \\
				\bottomrule
				\censorchange{}{\multicolumn{7}{@{}l}{\rule{0pt}{4ex}\small \emph{County names omitted for blind review.}}}
			\end{tabular}%
		}%
		\label{tab:significance-rtt-county}
	}
\end{table}

\boldparagraph{RTT per Country}
We first analyzed the RTT differences between countries. To do so, we calculated the median RTT per city to determine a stable average value for each of them. After that, we calculated the median RTT of each country's cities to determine their main tendencies. We assume that attackers may try to spoof the location via a VPN connection, so the measured RTT might not match their IP-indicated region. Thus, we considered successful logins separately, to compare the results at a higher level of trust, as it would be the case with RBA, i.e., only successful logins are stored%
.

Figure~\ref{fig:rtts_by_country} shows the results. Most countries had a higher median RTT than the online service's main country. Furthermore, the differences in the RTTs were significant for some countries (%
see Table~\ref{tab:significance-rtt-country}). All RTTs had a median standard deviation of 290~ms per country. Following that, the RTTs can be narrowed down to a number of countries.

\boldparagraph{RTT per Region}
Our data set contained a large number of RTTs with an IP geolocation inside the online service's home country. Thus, we further analyzed whether it is possible for the RTT to verify regions in this country. To consider a high level of trust, we only included successful logins%
. As in the previous study, we first calculated the median RTT per city. After that, we clustered the RTTs to each of the country's counties. The results show significant differences in the RTTs of the counties (%
see Table~\ref{tab:significance-rtt-county}). Following that, the RTT can also narrow down regions of a country.

\subsection{RTT as a IP Address Replacement}

Our previous results showed RTT's potential to verify IP-based geolocations. One step further would be to remove the full IP address from the feature set completely. Replacing the IP address with the RTT would protect users' privacy, e.g., in terms of data breaches~\cite{wiefling_privacy_2021}.

Based on the ideas of Wiefling et al.~\cite{wiefling_privacy_2021}, we only derived the ASN and country from the IP address, and then replaced the IP address with the RTT. The RTT had the identical weighting as the IP address for a fair comparison. \revisionadd{We kept and evaluated the RTT values as they were recorded each login attempt, to test the RBA behavior under real-world conditions.}

To compare the risk scores between IP address and RTT, we only included login history data that had both features, i.e., 2.5M legitimate logins. Based on the data set, we calculated the different risk scores for RBA when using IP address and RTT features. Similar to related work~\cite{wiefling_whats_2021}, we also rounded the RTT to the nearest five, ten, and 50 ms, to observe the RBA behavior at different levels of RTT granularity. We tested the naive, VPN, and targeted attacker. %
The available RTT data for the very targeted attacker was sparse, which is why we were not able to test this attacker here.

\begin{figure}[t]
    \centering
    \includegraphics[width=0.8\linewidth]{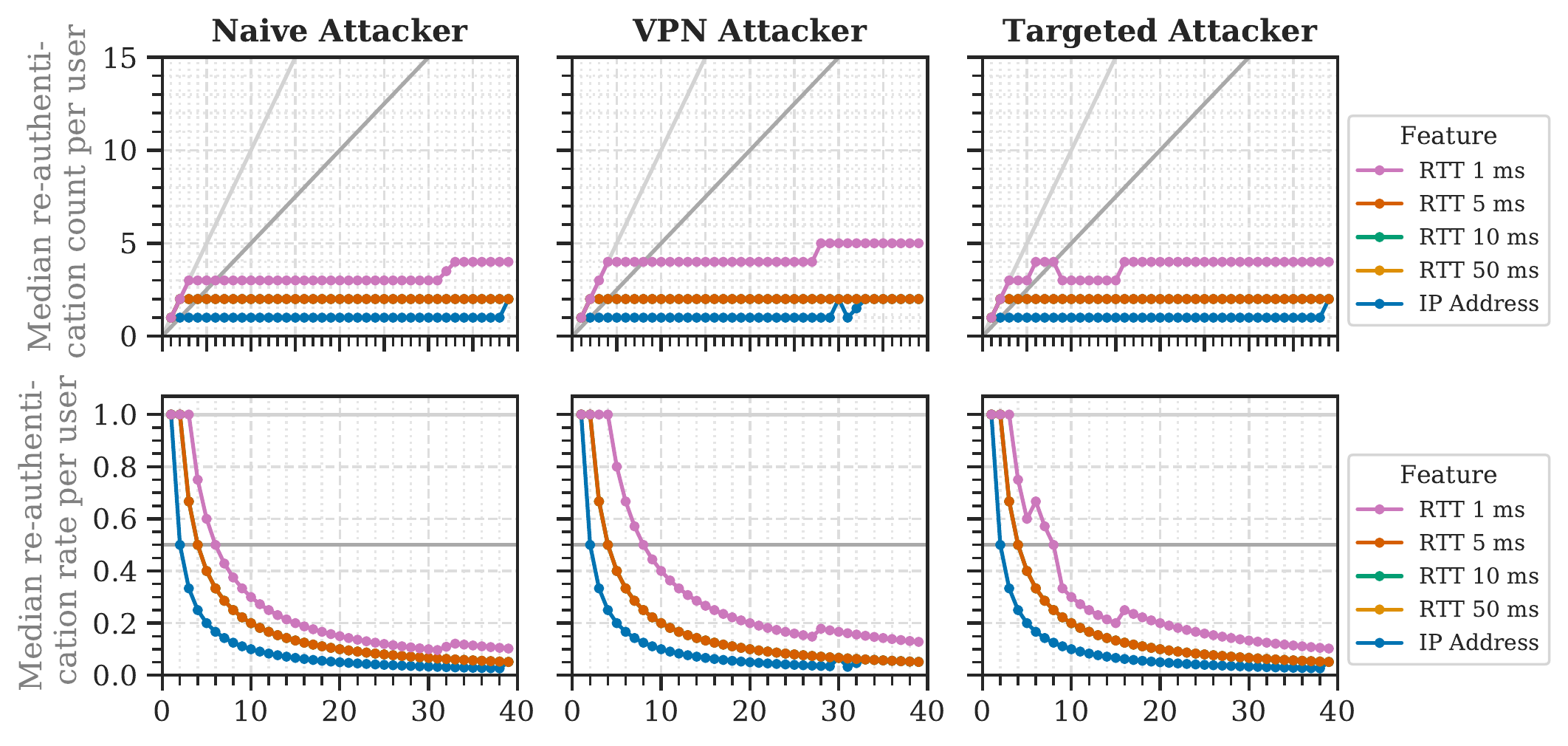}
    \caption{Direct comparison of the RTT and IP address features regarding their re-authentication counts and rates (TPR: 0.999). The results for 5ms, 10ms, and 50ms were identical.}
    \label{fig:re_auth_rtt_0.999}
\end{figure}

The median re-authentication rates of all features were zero until TPRs of 0.995. At a TPR of 0.999, the median re-authentication rates of the RTT slightly overtook those of the IP address for all attacker models (see Figure~\ref{fig:re_auth_rtt_0.999}).

To estimate the features' ability to distinguish between attackers and legitimate users, we also calculated the risk score relation (RSR)~\cite{wiefling_whats_2021,wiefling_privacy_2021} for them. The RSR is defined as the relation between the mean risk scores for attackers and legitimate users:
\begin{equation}
    RSR = \frac{mean\ attacker\ risk\ score}{mean\ legitimate\ risk\ score}
\end{equation}
The higher the RSR, the better attackers can be distinguished from legitimate users. Table~\ref{tab:rsr-rtt} shows the results for the features and attacker models. In general, the IP address achieved a slightly higher RSR than the RTT variations.

\begin{table}[t]
    \centering
    \caption{RSRs for the IP address and RTT features, based on the three attacker models (TPR: 0.999).}
    \begin{tabular}{@{}lrrrrr@{}}
    \toprule
    & & \multicolumn{4}{c}{RTT} \\
    \cmidrule{3-6}
    Attacker Model &    IP Address &  1 ms &  5 ms &  10 ms &  50 ms \\
    \midrule
        Targeted Attacker & 70.25 & 68.53 & 68.22 & 67.87 & 65.64 \\
               VPN Attacker & 61.90 & 60.39 & 60.12 & 59.81 & 57.83 \\
             Naive Attacker & 54.43 & 53.10 & 52.87 & 52.61 & 50.88 \\
    \bottomrule
    \end{tabular}%
    \label{tab:rsr-rtt}
\end{table}

\subsection{Discussion}
The results, especially those of successful logins, highlight the tendency that most RTTs were higher than those of the online service's main region. Therefore, the RTT can be considered a strong factor to verify locations indicated by the IP address, reflecting and extending previous findings~\cite{wiefling_whats_2021,rivera_risk-based_2020,abdou_secure_2018}. Beyond that, the RTT can also identify users when used as a replacement for the IP address, which is a new finding. \revisionadd{The RTT was measured only on mobile devices, so they contain larger variations due to mobile connectivity. We expect that these variations would be smaller on desktop devices connected to home or company networks. Therefore, it is possible that the study results are more negative than under real-world conditions involving desktop devices.}

The RTTs for Malaysia\censorchange{}{\footnote{\label{footnote:country-changed}Country name changed for blind review}} were very low. This is due to the fact that the online service provider operated mobile networks inside this country. Therefore, they were likely authenticated via their local mobile network. An optimized mobile connection was also likely true for Spain\censorchange{}{\footnotemark[\ref{footnote:country-changed}]}, where the RTTs were very low only in popular travel destinations for people of the online service's home country. \revisionadd{In practice, this corresponds to behavior similar to CDNs, so users inside these regions can be verified with high confidence.}

The IP address had higher RSRs, but the RSR differences to the RTT were low. The RSR differences between the 1 ms and 5 ms RTT variations were small. The latter, however, achieved a lower re-authentication rate. This rate was identical to those of 10 and 50 ms, which had lower RSRs. Therefore, a RTT of 5 ms provides good usability and privacy, and is preferable to 10 and 50 ms in terms of security in this data set and TPR context.

Using the RTT as a feature is getting more relevant, as some tech companies started testing VPN and proxy solutions as a privacy enhancement for their users~\cite{mozilla_vpn_2021,pauly_get_2021}. Considering a relatively stable delay between VPN server and online service, we assume that the RTT can also be used to determine the distance between a VPN server and the client device. This, however, requires further study.

For the RTT feature comparison, we had to recalculate the attacker models based on the reduced RTT data set. The majority of failed login attempts did not include a RTT. We expect that most attackers did not expect the mobile number authentication method to be successful and thus did not try it. We assume that this is why the re-authentication counts per TPR were lower than when using the full data set. Nevertheless, our results can still give indications on the effectiveness of using the RTT feature.

\section{Limitations}\label{sec:limitations}

The study results presented in this paper are representative for a certain country (Norway) and online service type (SSO of telecommunication services). They do not represent all online services worldwide, but show an example of a typical large-scale online service with sensitive data involved.

\revisionadd{Since the data set corresponds to the global browser market share (see Section~{\ref{sec:login-sessions}}), our results can help to apply RBA globally. Nevertheless, cultural differences in some countries may lead to different login patterns. For example, people in rural Myanmar rely on mobile shop staff to set up online accounts for them~{\cite{einzenberger_if_2016}}. In this case, online services can notice that many different user accounts were created over the same Internet connection. From a Western perspective, this behavior may seem suspicious, but it is normal in these areas. Therefore, it makes sense for service owners to tailor their RBA systems to cultural conditions in the target locations to achieve a high user acceptance.}

The RTT was measured only once per login attempt. We assume that the RSRs for the RTT would improve with multiple RTT measurements per login attempt, as shown in Wiefling et al.~\cite{wiefling_whats_2021} with five RTT measurements during the login process.

\section{Related Work}\label{sec:related-work}

With our contributions, we extend the related work in various aspects, which we discuss below.

Freeman et al.~\cite{freeman_who_2016} tested %
their RBA model in a case study using 300K legitimate logins collected during six months on the popular large-scale online service LinkedIn. However, they did not provide an analysis of RBA characteristics in practice. Also, for such a large-scale online service, the sample was rather small. To provide a more realistic estimate%
, we analyzed RBA characteristics using 12.5M legitimate logins collected in more than one year.
Wiefling et al.~\cite{wiefling_whats_2021} tested Freeman et al.'s model on a small online service with 9,555 legitimate logins collected over 1.8 years. Their and our results confirmed that IP address, user agent string, and RTT are useful RBA features. Our work also extended the analysis, e.g., for different login frequencies.
Wiefling et al.~\cite{wiefling_privacy_2021} proposed and tested privacy enhancements for RBA. Due to their small data set, they were not able to test their proposed login history minimization approach. We filled this research gap and tested this approach on a large-scale online service.

Alaca and van Oorschot~\cite{alaca_device_2016} evaluated a list of features regarding their ability to identify users. Regarding distinguishing information, they rated the IP address feature higher than the RTT. They noted, however, that the latter required further study.
There were different RTT implementations for RBA proposed in concurrent and independent work. Rivera et al.~\cite{rivera_risk-based_2020} proposed a client-originated version. Wiefling et al.~\cite{wiefling_whats_2021} proposed a server-initiated solution which they considered harder to spoof for attackers. The studies in both works showed the RTT's potential to verify countries~\cite{rivera_risk-based_2020} and users~\cite{wiefling_whats_2021}. Our work furthermore showed RTT's potential to verify regions, and to replace the IP address as a feature.
Campobasso and Allodi~\cite{campobasso_impersonation_2020} studied a criminal infrastructure that used malware-infected devices %
to bypass RBA. The RTT feature can detect such attempts, as the infrastructure used SOCKS5 proxies.

Doerfler et al.~\cite{doerfler_evaluating_2019} evaluated RBA re-authentication challenges on a sample of 1.2M legitimate Google users. They stated that their RBA implementation was able to block 99.99\% of automated account takeover attempts and 92\% of phishing attacks. Our study results with 3.3M legitimate users reflect their findings.

\revisionadd{Besides RBA, there are other related methods which aim to protect user accounts against automated attacks. CAPTCHAs~{\cite{von_ahn_captcha_2003}} are a widely used measure to block bot traffic by presenting a challenge that, in contrast to bots, are solvable by humans with a lower error rate. Some RBA systems also present CAPTCHAs in some occasions to increase friction only for bot-like traffic~{\cite{wiefling_is_2019}}. Solving CAPTCHAs, however, is still error prone for humans~}\cite{chalil_madathil_empirical_2019,reynaga_usability_2013}\revisionadd{, so achieving usable CAPTCHAs can be considered a challenge. Bots can also bypass them by proxying the CAPTCHA challenge to real humans~{\cite{wiefling_is_2019}} or machine learning algorithms~{\cite{sukhani_automating_2021}}. These bypass methods are scalable in practice. RBA, in contrast, depends on the specific user. Thus, proxying can not scale when RBA uses appropriate features, e.g., the RTT.}

\revisionadd{Another method is attribute-based authentication (attribute-based credentials, ABC)~{\cite{camenisch_concepts_2013}}. In this case, users authenticate to an online service by providing a verified attribute, e.g., their age to prove that they are over 18 years old. The verification is done by a trusted authority and the information presented to the online service can be done using pseudonyms. Therefore, online services only receive the information that is required for their service, e.g., that the user is over 18 without getting personal details. ABC can increase user privacy, but is independent from RBA, as the latter does not require verified personal details. However, ABC is thinkable as a re-authentication challenge in high risk environments where users expect increased security, e.g., online shopping or online banking~}{\cite{wiefling_more_2020,livraga_user_2016}}

\section{Conclusion \revisionadd{and Suggestions}}\label{sec:conclusion}

Online services are a popular attack target~\cite{akamai_loyalty_2020}. Stolen passwords used, for example, in credential stuffing attacks pose new security challenges to service developers, operators, and owners~\cite{doerfler_evaluating_2019}.
As these attacks increased for years~\cite{akamai_loyalty_2020,akamai_credential_2019}, RBA becomes an important measure for online services to protect their users. NIST, NCSC, and ACSC recommend RBA~\cite{grassi_digital_2017,national_cyber_security_centre_cloud_2018,australian_australian_2021}, some of them for more than three years.
Still, current literature does not provide any insights on implementing, configuring, and operating RBA on these types of online services. These insights are important, though, to achieve RBA's security gain while balancing usability and privacy. Only a fraction of work exists that focuses on RBA algorithms or studies, all of them on small online services. We studied RBA characteristics on a large-scale online service to close this research gap.

Our results, based on our data set of a large online service, confirm previous results, based on a small online service, that RBA rarely requests re-authentication in practice, even when blocking more than 99\% of targeted attackers. As new results, our evaluation furthermore shows that RBA can even detect a high number of very targeted account takeover attempts. The RBA behavior also significantly depends on the users' login frequencies.
For daily use websites, RBA can achieve very high security with few re-authentication requests. For monthly use websites, the percentage of blocked attackers needs to be lowered down to around 2-3\% to achieve a similar usability for targeted attackers. Still, a high percentage of attackers would be blocked for all user types. Therefore, we consider RBA an effective measure to protect users from the majority of attacks involving stolen passwords.
Besides that, the RTT showed to be a promising RBA feature to verify regions and to identify users while increasing user privacy.

RBA needs to be carefully evaluated for and adapted to each individual use case scenario. Based on our findings, RBA system designers should consider the following to find the best configuration for their environment:
\begin{enumerate*}
    \item The amount of minimizing the login history needs to be carefully adjusted to users' login frequencies. Online services with a daily user base should keep at least six months of login history for a stable setup that blocks targeted attackers. When blocking naive attackers only, which represented 97\% of all identified incidents at the online service, reducing to three months is possible.
    \item Attack data should be used only with caution or not at all, as these can greatly decrease the usability and security properties.
    \item To achieve an optimal basic configuration in short time, administrators can use the introduced ML-based RBA parameter optimization. For further adjustment, they can increase or reduce the resulting access thresholds to decrease or increase the security and average re-authentication count, respectively.
    \item Non-optimized RBA is potentially vulnerable to denial of service when multiple users log in at similar times. To mitigate this, and to considerably shorten the authentication time, the implementation of the RBA algorithm needs to be carefully designed using efficient data structures such as hash tables.
\end{enumerate*}

Our results and the synthesized data set support online services to gain a widespread understanding of RBA. In future work, we will use these insights to create an open source RBA solution that significantly improves usability, security, and privacy compared to the only open solution currently available, i.e., OpenAM~\cite{openam_adaptive_2016}. The data set will be used as testing data to evaluate our solution.

\begin{acks}
We would like to thank Rudolf Berrendorf and Javed Razzaq for providing us a huge amount of computational power for our big data analysis. Also, thanks to Florian Dehling and Jan Tolsdorf for providing feedback on the paper. 
This research was supported by the \grantnum{nerdnrw}{research training group ``Human Centered Systems Security'' (NERD.NRW)} sponsored by the \grantsponsor{nerdnrw}{state of North Rhine-Westphalia}~.
The Platform for Scientific Computing was supported by the \grantsponsor{13FH156IN6}{German Ministry for Education and Research}~, and the \grantsponsor{13FH156IN6}{Ministry for Culture and Science of the state North Rhine-Westphalia}~ (research grant \grantnum{13FH156IN6}{13FH156IN6}).
\end{acks} 

\bibliographystyle{ACM-Reference-Format}
\bibliography{bibliography}

\appendix

\newpage
\appendix

\section{Synthesized Data Set}\label{appendix:synthesized-data-set}

Our synthesized data set can be downloaded at \url{https://github.com/das-group/rba-dataset}.
As this paper introduces our open data set, we ask researchers to cite this paper whenever they publish own results obtained from adopting our data set. 
For broad compatibility, we provide the data as a CSV file. Table~\ref{tab:synthesized-features} shows the available features in the synthesized data set and the data types. We kept the feature values in readable notation to facilitate understanding. The values may need to be converted to more efficient data types to speed up risk score calculation. For instance, the IP address could be converted to its integer value. Other string-based features could be hashed and converted to integer values as well.

Common synthetic data generation methods rely on learning the real values of a data set, which, however, does not provide a strong privacy guarantee~\cite{stadler_synthetic_2021}. Therefore, we used our own generation method that was tailored to the use in RBA models used by the majority of online services in the wild~\cite{freeman_who_2016,openam_adaptive_2016,wiefling_whats_2021}. %
We created the synthetic data set as follows: As common RBA models only rely on the number of occurrences of categorical data, we did not have to reuse any of the original feature values. We transformed all unique data set values to a random pseudonymous identifier and added randomness to the timely order, to keep the statistical relations between different feature values. Then, we randomly assigned feature values from a Location-to-ASN-to-IP data set in backwards hierarchy (i.e., Country $\rightarrow$ Region $\rightarrow$ City $\rightarrow$ ASN $\rightarrow$ IP address). If the original IP address was used in an attack, we made sure that the randomly assigned IP address is an attack IP address as well. We repeated the same procedure for the user agent strings. Depending on the synthesized geolocation data and the login success status, we randomly drew the RTTs for the corresponding entries, to ensure that the RTT results are realistic ones. %
Finally, we generated the login timestamps with random noise following the distributions and the timely order (see Figure~\ref{fig:login_attempts_per_day_and_minute} in Section~\ref{sec:performance}).

We verified the synthesized data set by recalculating the study results from this paper. As our generation process does not reuse feature values from the original data set, the synthesized data set should not provide any personal identifiable information. Nevertheless, we evaluated the privacy attributes of the synthesized data set using Stadler et al.~\cite{stadler_privacy_2021} and other comparable frameworks.

\begin{table}[H]
	\centering
	\caption{Available features in the synthesized data set}
	\resizebox{0.75\linewidth}{!}{%
		\begin{tabular}{@{}lll@{}}
			\toprule
			Feature & Data Type & Range or Example \\
			\midrule
			IP Address & String & 0.0.0.0 - 255.255.255.255 \\
			Country & String & US \\
			Region & String & New York \\
			City  & String & Rochester \\
			ASN   & Integer & 0 - 65535 \\
			\midrule
			User Agent String & String & Mozilla/5.0 (Windows NT 10.0; Win64; ... \\
			OS Name + Version & String & Windows 10 \\
			Browser Name + Version & String & Chrome 70.0.3538 \\
			Device Type & String & (mobile, desktop, tablet, bot, unknown) \\
			\midrule
			User ID & Integer & [Random pseudonym] \\
			Login Timestamp & Integer & [64 Bit timestamp] \\
			RTT [ms]  & Integer & 1 - 8600000 \\
			\midrule
			Login Successful & Boolean & (true, false) \\
			Is Attack IP & Boolean & (true, false) \\
			Is Account Takeover & Boolean & (true, false) \\
			\bottomrule
		\end{tabular}%
	}
	\label{tab:synthesized-features}%
\end{table}%

\onecolumn

\section{Full Results: Evaluating RBA in Practice (RQ1)}

\begin{figure}[H]
    \centering
    \includegraphics[width=\linewidth]{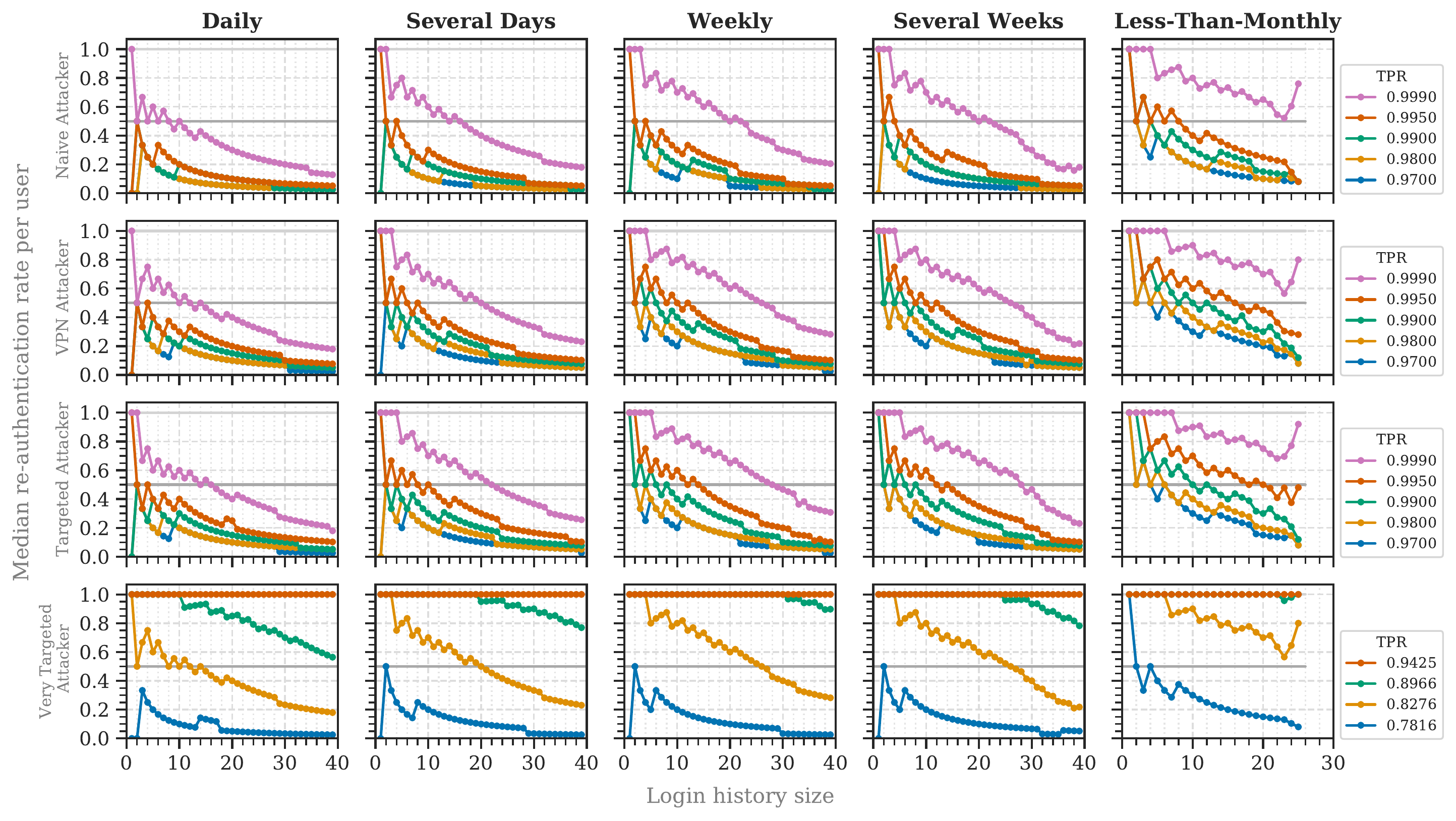}
    \caption{Median re-authentication rates based on the login frequency and attacker model}
    \label{fig:average_re_authentication_rates_usage_full}
\end{figure}

\section{Full Results: Login History Minimization (RQ3)}

\begin{figure}[H]
    \vspace{-1em}
    \centering
    \includegraphics[width=\linewidth]{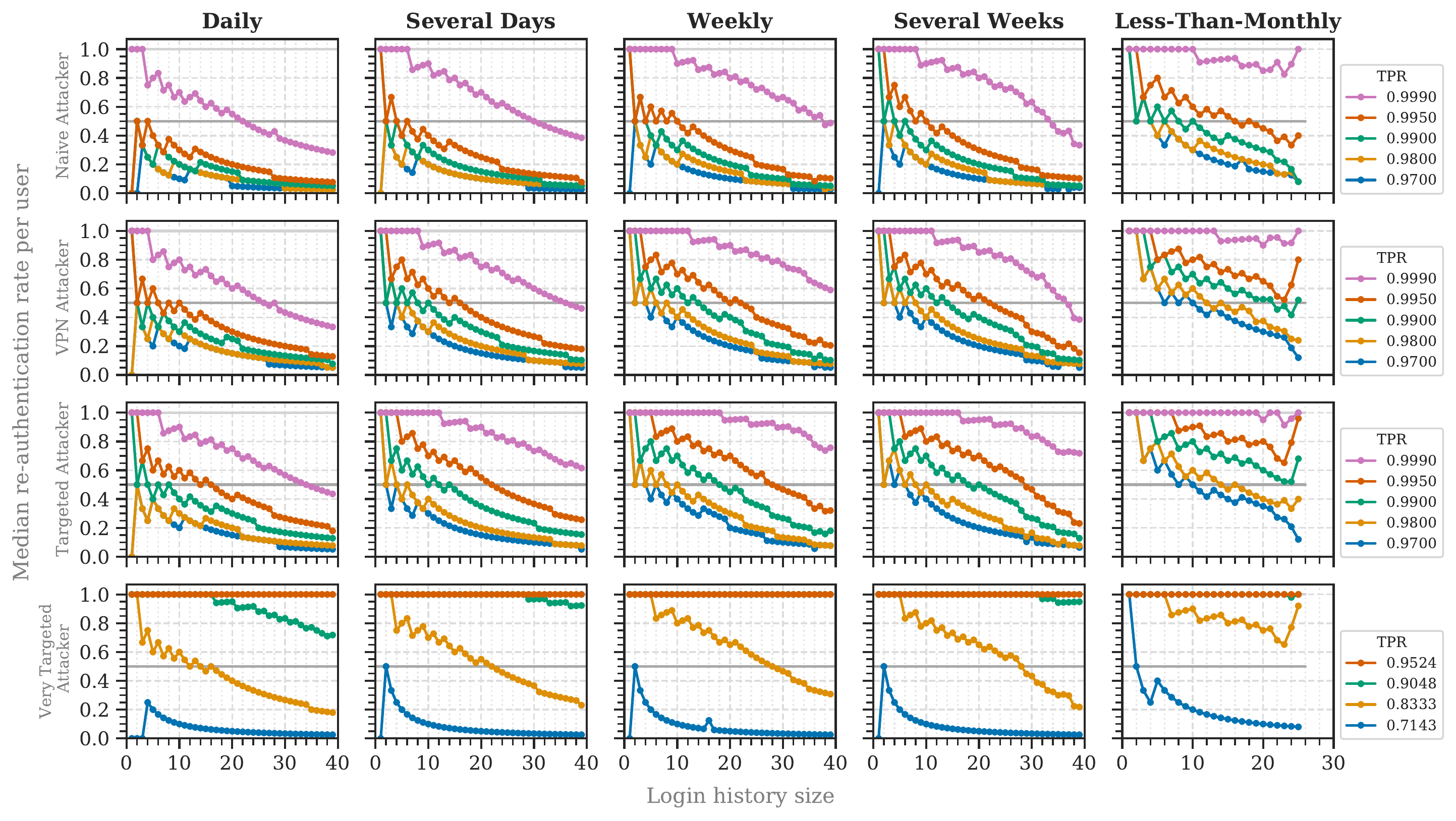}
    \caption{Median re-authentication rates based on the login frequency and attacker model (minimized 6 months)}
    \label{fig:average_re_authentication_rates_usage_full-6_months}
\end{figure}

\begin{figure}[H]
    \centering
    \includegraphics[width=\linewidth]{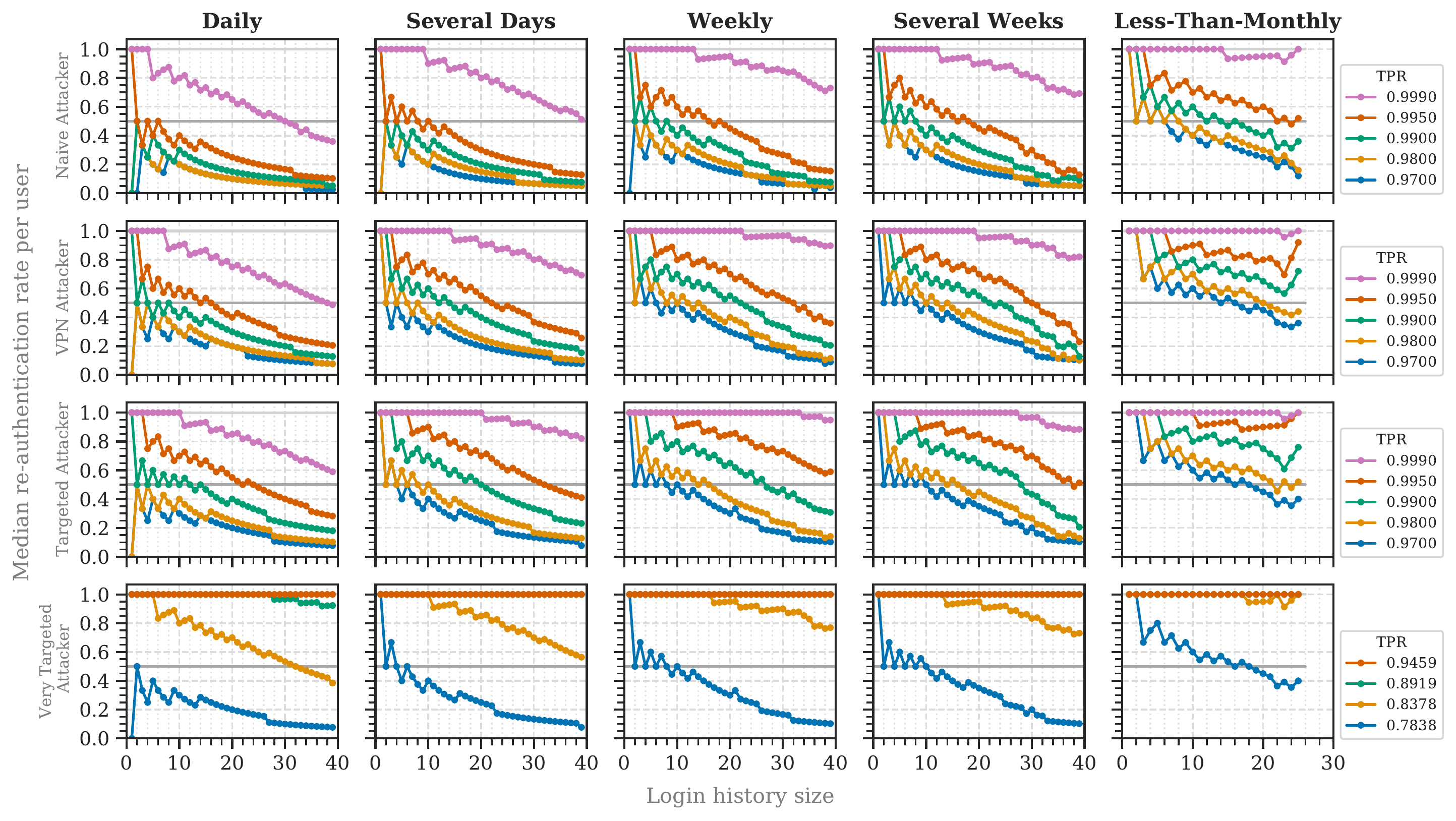}
    \caption{Median re-authentication rates based on the login frequency and attacker model (minimized 3 months)}
    \label{fig:average_re_authentication_rates_usage_full-3_months}
\end{figure}

\begin{figure}[H]
    \centering
    \includegraphics[width=\linewidth]{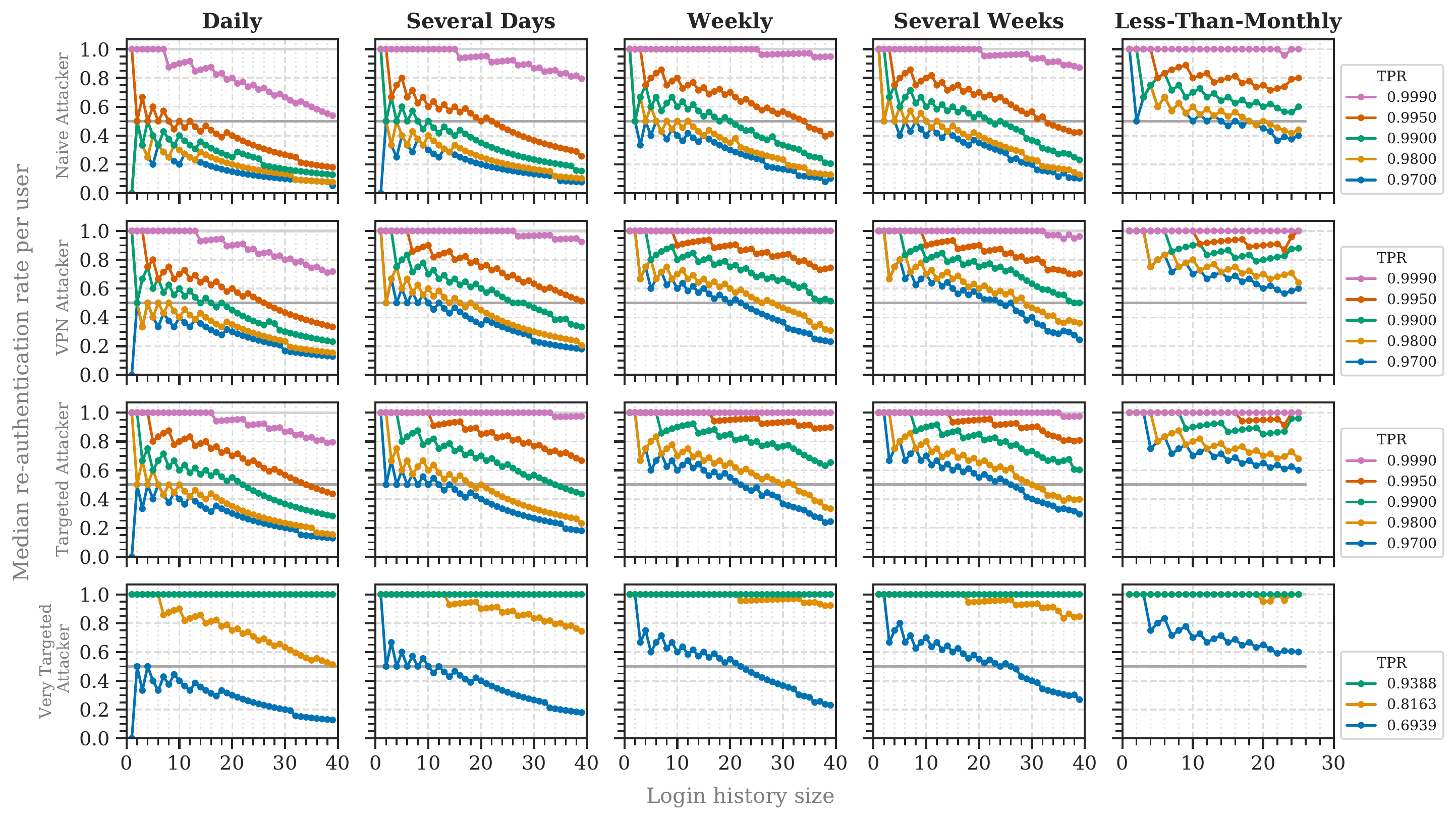}
    \caption{Median re-authentication rates based on the login frequency and attacker model (minimized 1 month)}
    \label{fig:average_re_authentication_rates_usage_full-1_months}
\end{figure}

\newpage

\section{ML-Based RBA Parameter Optimization (RQ4)} \label{appendix:ml-results}

\begin{table}[H]
    \centering
    \caption{RQ4, Study 1: Achieved TPRs for the ML-generated dynamic access thresholds in practice. We trained the ML models on 24 consecutive slices of the data set. The hybrid method achieved higher TPRs than the linear and poly models.}
    \resizebox{0.35\linewidth}{!}{%
    \begin{tabular}{@{}llll@{}}
    \toprule
    ML Model &     Range     & Median    & SD    \\
    \midrule
    \multicolumn{4}{@{}l}{\emph{Naive Attacker}}\\
    Linear    &  0.9319-0.9961 &  0.9713 &  0.0173 \\
    Poly      &  0.9312-0.9966 &  0.9718 &  0.0179 \\
    Hybrid    &  0.9338-0.9967 &  0.9730 &  0.0170 \\
    \midrule
    \multicolumn{4}{@{}l}{\emph{VPN Attacker}}\\
    Linear    &  0.8646-0.9929 &  0.9250 &  0.0319 \\
    Poly      &  0.8640-0.9954 &  0.9249 &  0.0335 \\
    Hybrid    &  0.8671-0.9957 &  0.9278 &  0.0327 \\
    \midrule
    \multicolumn{4}{@{}l}{\emph{Targeted Attacker}}\\
    Linear    &  0.8646-0.9929 &  0.9283 &  0.0324 \\
    Poly      &  0.8646-0.9949 &  0.9290 &  0.0338 \\
    Hybrid    &  0.8670-0.9951 &  0.9316 &  0.0331 \\
    \midrule
    \multicolumn{4}{@{}l}{\emph{Very Targeted Attacker}}\\
    Linear    &  0.6782-0.8506 &  0.7701 &  0.0520 \\
    Poly      &  0.6552-0.8621 &  0.7931 &  0.0523 \\
    Hybrid    &  0.6782-0.8736 &  0.7931 &  0.0500 \\
    \bottomrule
    \end{tabular}%
    }
    \label{tab:dynamic-access-thresholds-tprs}
\end{table}

\section{SIMPLE Model Evaluation} \label{appendix:simple-model}

\begin{figure*}[h]
    \centering
    \revisionaddimage{\includegraphics[width=0.98\linewidth]{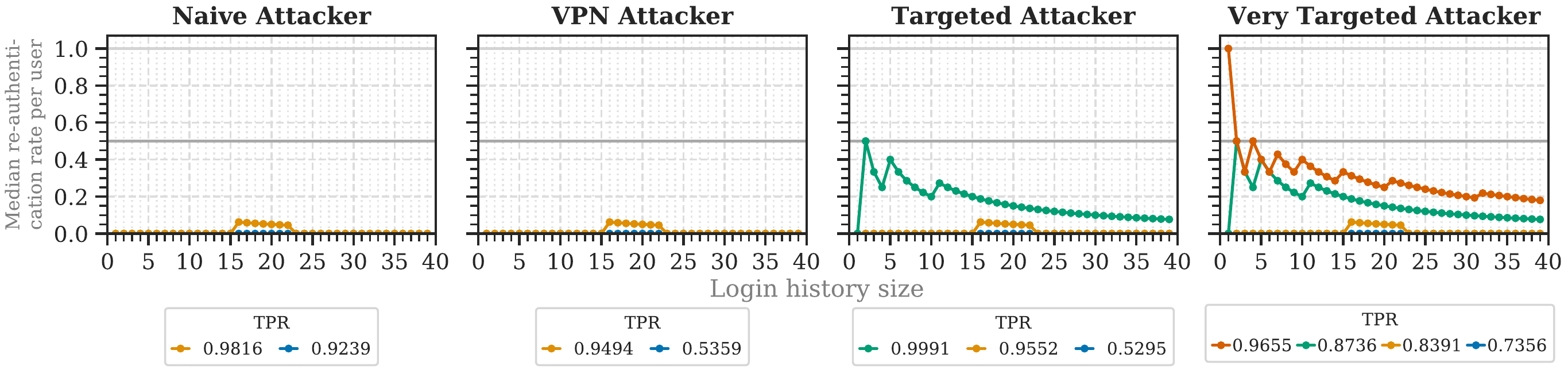}}
    \caption{\revisionadd{Median re-authentication rates for the four attacker models when using the SIMPLE model. The model's risk score is coarse-grained. The TPRs therefore show the configuration steps that are possible under these conditions. For orientation, we added the baseline for 2FA (light grey line), and the stable setup threshold (grey line).}}
    \label{fig:average_re_authentication_count_rates_combined-simple}
\end{figure*}

\revisionadd{To provide a baseline evaluation, we repeated the study of RQ1a and RQ1c (see Section~{\ref{sec:testing-rba}}) with the SIMPLE model used in the open source SSO solution OpenAM~{\cite{openam_adaptive_2016}} (see Section~{\ref{sec:rba-model-selection}}). The model checks the feature values for an exact match in the user's login history. In contrast to the Freeman et al. model{~\cite{freeman_who_2016}}, the SIMPLE model does not allow to take attack data into account. As in the other studies, the model used the IP address and user agent string as features. The results show that the achievable TPRs were very coarse-grained, e.g., the TPR for targeted attackers dropped from 0.9552 to 0.5295 when the access threshold was lowered by a tiny fraction to the next possible level (see Figure~{\ref{fig:average_re_authentication_count_rates_combined-simple}}). This confirms previous findings~{\cite{wiefling_whats_2021}}. Also, the high end of TPRs varied largely across the different attackers, e.g., 0.9816 for naive attackers and 0.9494 for VPN attackers. This uncertainty makes the SIMPLE model difficult to configure for large-scale online services.}

\end{document}